# High Precision Orientation Mapping from 4D-STEM Precession Electron Diffraction data through Quantitative Analysis of Diffracted Intensities

**Leonardo M. Côrrea[1], Eduardo Ortega[2], Arturo Ponce[2], Mônica A. Cotta[1], Daniel Ugarte[1,*]**

1. **Instituto de Fisica "Gleb Wataghin", Universidade Estadual de Campinas-UNICAMP, 13083-859, Campinas - SP, Brazil**
2. **Department of Physics and Astronomy, University of Texas, San Antonio, Texas 78249, United States**

* Email: dmugarte@ifi.unicamp.br; Number: 55-19-35215384;

## Abstract

The association of scanning transmission electron microscopy (STEM) and the detection of a diffraction pattern at each probe position (so-called 4D-STEM) represents one of the most promising approaches to analyze structural properties of materials with nanometric resolution and low irradiation levels. This is widely used for texture analysis of material using automated crystal orientation mapping (ACOM). Herein, we perform orientation mapping in InP nanowires exploiting precession electron diffraction (PED) patterns acquired by an axial CMOS camera. Crystal orientation is determined at each probe position by the quantitative analysis of diffracted intensities minimizing a residue comparing experiments and simulations in analogy to structural refinement. Our simulations are based on the two-beam dynamical diffraction approximation and yield a high angular precision (~0.03º), much lower than the traditional ACOM based on pattern matching algorithms (~1º). We anticipate that simultaneous exploration of both spot positions and high precision crystal misorientation will allow the exploration of the whole potentiality provided by PED-based 4D-STEM for the characterization of deformation fields in nanomaterials.

# 1. Introduction

The possibility of controlling chemical or physical properties of nanostructured materials have transformed then into essential constituents of many technological devices, such as sensors, electromagnetic or optical devices, *etc.* (Lieber, 2003). For example, semiconductor nanowires (NWs) are now proposed for many optoelectronic systems, where the application performance can be optimized by modifying their size, chemical composition, atomic structure, structural defects, *etc.* (Jia et al., 2019). The presence of defects can affect their properties in unexpected ways, but usually inducing detrimental performance. Although the most common synthesis methods can provide high-quality crystalline order, the nanometric dimensions of the NW can favor the formation of metastable structure and defects/distortion may be unavoidable (Güniat et al., 2019). The control of such aspects require precise analytical tools, which display unique challenges for such small systems.

Transmission electron microscopes (TEMs) are widely used in the characterization of nanomaterials because of the achievable high spatial resolution. The huge recent developments in instrumentation aspects (aberration correctors, direct detection cameras, *etc.*) provides structural information at the atomic scale, which is considered accessible and reliable (Thomas et al., 2015; Willians & Carter, 2009). In general, advanced TEM methods are associated with atomic resolution imaging, but recently the development of the so-called four-dimensional scanning-TEM (4D-STEM) have raised the interest in electron diffraction techniques (ED) (Ophus, 2019). In this technique, a focused electron beam is scanned over the sample and a ED, or scattering pattern, is record for each pixel position. The resulting data can be utilized for several quantitative analysis methodologies to measure the structural properties of materials sample (phase, strain, thickness, *etc.*) with nm resolution. From the local ED pattern, many different data types can be derived such as virtual detector imaging (VBF: virtual bright field, VDF: virtual dark field, VADF: virtual annular dark field), higher-order Laue zone analysis, *etc.* (Ophus, 2019). Overall, 4D-STEM allows for versatile data acquisition and precise structural analysis without some of the limitations of atomic resolution imaging techniques, in particular reducing electron dose, allowing larger fields of view, *etc.*

Additionally, recently 4D-STEM has been applied to further developments in the collection and analysis of big data to extract the transmitted electron wave phase (lost during detection), in which the phase is encoded throughout a diffraction pattern at large convergence angle (Li et al., 2022; Ophus, 2019). Alternatively, 4D-STEM analog analyses have been applied in the Fourier space of atomic resolution images and applied to crystal segmentation of multiple twinned small particles (Bárcane-González et al., 2022) .

The fundamental and tricky problem associated with electron diffraction interpretation is the phenomenon described as dynamical diffraction. Due to the very strong interaction between the electron beam and the sample, the electrons are scattered many times within the specimen, leading to complex effect on diffracted beam intensities, nonhomogeneous illumination of diffracted spots, *etc*. (Willians & Carter, 2009). The prediction of dynamical diffraction effects require rather slow and complex numerical simulations (Kirkland, 2020). An interesting alternative approach is the so-called precession electron diffraction (PED), where a pattern is generated by the incoherently addition of EDs acquired while the electron beam precesses around the microscope optical axis (forming a hollow cone) (Vicent & Midgley, 1994). The resulting PED pattern display diffracted beam intensities with reduced dynamical diffraction effects; this regime is called quasi-kinematical and greatly improves the quantitative interpretability of diffraction patterns (Midgley & Eggeman, 2015; Oleynikov et al., 2007; Own et al., 2006; Rauch et al., 2010; Vincent & Midgley, 1994). In addition, diffraction disks show more homogeneous intensity, and more reflection are excited in high-scattering angles; this last attribute is fundamental to obtain information about small interatomic distances.

Scanning precession electron diffraction (SPED, can be included as a 4D-STEM method) is also widely used in materials science for texture analysis by means of the so-called automated crystal orientation mapping (ACOM) (Rauch et al., 2010; Viladot et al., 2013). In fact, orientation indexing of each PED pattern can be carried out automatically by comparison with a library of kinematically calculated ED patterns. ACOM studies are usually performed in large areas, such that up to tens of thousands of EDs can be measured in a single experiment. Traditionally, TEM detectors are based on the charge coupled devices (CCD) technology, were usual read-out speeds are in order of ~1 frame/s, which are unsuited for ACOM (*e.g.* 10000 EDs imply in ~ 3h of measurement) (MacLaren et al., 2020). For that reason, the conventional PED set-up uses a high-speed external optical camera to capture diffraction patterns from the TEM fluorescent screen during nanobeam scanning (Moeck et al., 2011). The external optical camera captures the SPED patterns as images with an off-axis geometry, which must follow additional postprocessing steps, to correct different severe image distortions (Moeck et al., 2011). Although a high-frame rate is possible, acquired patterns contain afterimages from the last several probe positions because the fluorescent screen continues to emit light (~100 ms) after electron impact (Moeck et al., 2011). In addition, due to the various electron to light and light to electron conversion steps, the signal-to-noise ratio (SNR) for this technique is suboptimal. Hence, this traditional SPED setup may introduce artifacts, obscure the detection of weak reflections and strongly reduces angular resolution.

A much better measurement of PED patterns could be realized by using an on-axis geometry camera to reduce geometrical distortions. The required high acquisition speed may be fulfilled by complementary metal-oxide-semiconductor (CMOS) detectors based on monolithic active pixel sensors (APS). This CMOS cameras allows bigger arrays, faster readout (similar to the off-axis optical CCD), higher-dynamical range, reduced noise, and reduced sensitivity to electron radiation damage (Jeong et al., 2021). More importantly, new opportunities for analysis methods are possible as those detectors preserve the linearity of the intensities. The conventional ACOM methods relies on a template-matching procedure where orientation identification is performed through a cross-correlation between the measured ED and a library of simulated patterns (see Figure 1). This numerical calculation mostly takes into account diffraction peaks position, and intensities are usually rescaled. For example, in order to enhance angular resolution, high angle scattered beams are rescaled by gamma function or even binarized (Jeong et al., 2021). Some authors have applied the Laue circle method to fit diffraction spot positions and quickly determine sample disorientation (see Figure 1) (Ben-Moshe et al., 2021; Edington, 1975).

The main strength of x-ray diffraction techniques (probably the most popular structural characterization method) is the quantitative use of the diffracted intensities though the comparison between measurement and simulations by mean of residue analysis (Billinge & Levin, 2007). In recent years, PED data has played a fundamental role to consolidate electron crystallography, allowing crystal structure solution/refinement analysis (as in x-ray or neutron diffraction) (Midgley & Eggeman, 2015). A PED pattern intensity distribution carries precise information about the orientation of the studied crystal, and the angular position of frames in electron diffraction tomography experiments have already been recalibrated using diffracted beam intensity analysis with an angular precision up to ~ 0.1° (Palatinus et al., 2019). The precedent precision is much less than for common ACOM methods, where angular precision is usually accepted to be around ~ 1° (Cautaerts et al., 2022; Harrison et al., 2022; Jeong et al., 2021; Ophus et al., 2022; Ugarte et al., 2019).

Herein, we demonstrate orientation mapping with PED-assisted 4D-STEM using a scintillator-coupled CMOS detector in TEM. For each nanobeam probe position, a high-quality PED pattern is acquired and the crystal orientation is automatically derived at each pixel by the quantitative analysis of diffracted intensities. We have studied semiconductor InP NWs displaying an axial screw dislocation and the corresponding crystal torsion (or Eshelby twist) (Eshelby, 1953; Eshelby, 1958). Our results indicate that it has been possible to achieve much higher angular precision (~ 0.03º), well below normal ACOM method reports (~ 1º) and structure refinement efforts (~ 0.1º) (Cautaerts et al., 2022; Harrison et al., 2022; Jeong et al., 2021; Ophus et al., 2022; Palatinus et al., 2019; Ugarte et al., 2019).  We show a detailed comparison between different ACOM methodologies and make a rigorous analysis of the results quality and reliability. This precision improvement has been fundamental to observe the small crystal disorientation in the Eshelby twisted NWs.

# 2. Materials and Methods

## 2.1 Sample and measurements.

The In(Ga)P nanowires utilized in this work were grown using the vapor-liquid-solid (VLS) mechanism using a Au catalysts inside chemical beam epitaxy (CBE) process. The generation of the wires has already been described in detail in (Ugarte et al., 2019) and (Tizei et al., 2011). Briefly, thermally decomposed phosphine ($PH_3$), trimethyl Indium (TMI) and triethyl Gasidellium (TEGa) have been used as precursors, with $H_2$ as carrier gas. The electron microscopy samples have been generated by gently scrapping the carbon grid (lacey type) on the substrate. The atomic arrangement in the wires is a wurtzite structure ($P6_3mc,\ a = b = 0.4150\ nm,\ c = 0.6912\ nm$); this hexagonal phase is not stable in macroscopic InP crystals.

The NWs represent a rather complex system to be analyzed (see Figure 2); the wires contain an axial screw dislocation (wire axis along wurtzite hexagonal $c$ - axis) such that the basal planes in the crystal form a helicoid of pitch $B$ (dislocation Burgers vector) (Eshelby, 1953; Eshelby, 1958). The normal to the helicoidal basal planes are tilted in relation to the wire geometrical axis and the tilt display opposite sense at different sides of the dislocation line. In addition, the torque created by the screw dislocation induces a backwards torsion to get mechanical stability for a finite volume, the so-called Eshelby twist, which was initially proposed in the 50s using classical elastic theory (Eshelby, 1953; Eshelby, 1958). The twist rate $\alpha$ (given as radians per nm) is given by:

$$\alpha = \frac{B}{\pi R_w^2} \quad (1)$$

where $B$ is the norm of the dislocation Burgers vector (in this case, $B = c$) and $R_w$ is the NW radius. The twist should easily be revealed as a continuous rotation of the crystal normal along the wire. This effect has been first observed in 2008 in PbSe and PbS NWs (Bierman et al., 2008; Zhu et al., 2008). A detailed description of Eshelby torsion in InP wires has been recently present by Ugarte *et al.*, (Ugarte et al., 2019) and the combination of crystal torsion superimposed to helicoidal basal planes represent quite challenging case to test new ACOM methodologies.

TEM images and electron diffraction patterns have been acquired on a JEOL 2010F microscope at 200 kV. Diffraction patterns were recorded in a highly sensitive 16-megapixel F416 CMOS camera (TVIPS) and they were registered at 16-bit to ensure maximum dynamical range. Subsequently, the patterns have been binned to 512x512 pixels. A Nanomegas PED unit was used to generate the precessing electron beam (frequency 100 Hz, precession angles 8.8 mrad). The beam full convergence angle was 1.84 mrad (measured from diffraction spot diameter), generating an approximately 5 - 8 nm diameter electron probe.

## 2.2 Data reduction.

Data treatment has been performed with a home-made software developed in Python and, all the steps (from measurement, data reduction and quantitative analysis) are described below or in a more detailed Supplementary Information. A key point in our procedure is that the ACOM results are derived from structural refinement based on the analysis of diffraction intensities.

The original data was acquired as movie (time resolved) as the detection CMOS axial camera was not synchronized with the PED attachment used in our experiments. To convert the diffraction pattern series into 4D data, we have used VADF intensity from each pattern to assign a spatial position and build the NW VADF image (see Figures 2 and 3). Once the 4D data set was generated a cross-correlation procedure was applied to detect diffraction spots position for indexation and also for the measurement of diffracted intensities from each diffraction disk (see Figure 4 and details in Supplementary Information). Peaks which have intensity below a defined threshold have not been included as their information can be significantly affected by noise (see Supplementary Information). This study aims the exploration of 4D-STEM PED intensities, so no attempt has been done to detect strain though the geometrical analysis of disk positions in the diffraction pattern (Cooper et al., 2016; Grieb et al., 2018; Han et al., 2018).

## 2.3 Diffracted Intensity Calculation

The measured PED intensities include both dynamical diffraction effects (strongly depend on crystal orientation and sample thickness) for each incident electron beam direction and, the incoherent integration along the beam precession circles. Due to the large number of diffracted patterns on an orientation mapping data set, we have chosen to use dynamical two-beam approximation for intensity calculation and a numerical integration to consider the change of excitation error associated to beam precession (as suggested by Gjonnes, Own and Oleynikov) (Gjonnes, 1997, Oleynikov et al., 2007; Own, 2005). Our rather small 4D data-set includes more than 3000 measured EDs, and each one may need to be simulated many times during the residue ($R_N$) minimization procedure in order to perform the ACOM analysis or crystal orientation optimization by quantitative comparison of diffracted intensities. Although the two-beam expression may be considered a rather simplified approach to dynamical diffraction when compared with multi-slice calculation and Bloch wave formalism, it yields an analytical expression which can be utilized for quick calculation of diffracted intensity (Willians & Carter, 2009). The two-beam approximation is considered of limited application to describe dynamical electron diffraction intensities from crystal oriented along a zone axis (strong many beam condition) (Willians & Carter, 2009). However, two-beam can provide good numerical descriptions of PED intensities, as beam precession reduces multiple scattering effects, as already reported by many studies (Palatinus et al., 2015; Sinkler et al., 2007).

The two-beam model (Howie-Whelan equations) gives the most basic description of the diffracted intensities ($I_{hkl}$) in the dynamical regime: (Willians & Carter, 2009)

$$I_{hkl} = \frac{\pi^2}{\xi_{hkl}^2} \frac{sin^2(\pi t s_{eff})}{(\pi s_{eff})^2} \quad (2)$$

where $s_{eff} = \sqrt{s^2 + \xi_{hk}{}^{-2}}$ is the effective excitation error, $t$ is the crystal thickness and $s$ is the excitation error for a given disorientation; $\xi_{hkl}$ is the extinction distance and it is defined as $\xi_{hk} =$ $\pi V_c/\lambda F_{hkl}$, where $V_c$ is the volume of the primitive cell of the crystal, $\lambda$ the electron wavelength and $F_{hkl}$ the structure factor (Reimer & Kohl, 2008). The application of this model requires knowledge of thickness of the analyzed region and precise orientation (though excitation error) to calculated diffraction peaks intensities.

Our simulations uses a numerical integration to consider the effects of beam precession geometry where spot intensity is integrated around the relaxed Bragg condition (integrating a range of $s$ values). We have calculated this by considering the Laue circle center ($O_L$) movement, as the electron beam precession implies in a complete revolution of $O_L$ around the transmitted beam position ($T$) (Gjonnes, 1997; Own, 2005). If $\theta$ is the angle that defines projection of the scattering vector $\boldsymbol{k}$ ($k = 1/\lambda$) (azimuthal position of the precessing beam, see Figure S3) we can express the PED intensity as: (Gjonnes, 1997; Own, 2005)

$$I_{hkl}^{PED} = \int_{0}^{2\pi} I_{hk}\,(\theta)d\theta \quad (3)$$

The precession geometry implies in a complex relation between the crystal orientation, precession angle and the excitation error. Such relation can be expressed by the equation derived by Gjonnes: (Gjonnes, 1997)

$$s(\theta) = -\frac{2R_0 g_{xy}\cos(\theta) + 2kg_z + g^2}{2k} \quad (4)$$

where $R_0$ is the Laue circle radius (${R_0}^2 = {k_x}^2 + {k_y}^2$), $\boldsymbol{g}$ is a point in reciprocal space, $g_{xy}$ is its component in the plane perpendicular the component $g_z$ parallel to the optical axis. Usually, the $g_z$ in Equation 4 is neglected due to the low value in the first Laue zone (most spots appear in this region due to the small curvature of the Ewald sphere for high energy electrons) (Palatinus et al., 2019; Sinkler et al., 2007; Vicent & Midgley, 1994). Although this may be a sound argument for analyzing zone axis patterns, its implementation forces a symmetry to the diffraction intensities (not suitable for simulations targeting ACOM). For the pair of reflections $hkl$ and $\bar{h}\bar{k}\bar{l}$, $g_{xy}$ will be identical and both reflections will have equal intensities and independent of crystal orientation (Hammond, 2015). Then, it is fundamental to include the $g_z$ term, as tiny value modification may induce asymmetry between intensities which carry the targeted essential information about crystal orientation.

In the following section we will describe the crystalline structure and crystal orientation referential implement for PED intensity calculation based on the two-beam approximation.

## 2.4 <u>PED intensity calculation and residue minimization.</u>

We have chosen our coordinate system such that initially the crystal is perfectly oriented at a zone axis set along the $x$ axis (anti-parallel to the microscope optical axis). Subsequently, the reciprocal lattice vector $\boldsymbol{g}$ of a reflection in a twisted crystal relates to the one in the ideal zone axis orientation ( $\boldsymbol{g_0}$) by $R_y(\rho)R_z(\beta)\,\boldsymbol{g_0}$, where $R_y$ and $R_z$ are the rotations matrix around the $y$ and $z$ axis, respectively. This issued to calculate the excitation error as a function of $\beta$, $\rho$ and $\theta$ ($\theta$ being the azimuthal position of the precessing beam) for any orientation of the crystal near the considered initial zone axis.

The next step involves the calculation of the structure factor for the estimation of $\xi_{hkl}$. In the case of InP wurtzite structure, two Debye-Waller factors ($B_{In}$ and $B_P$, for the In and P atoms respectively) are necessary. In our residue minimization procedure, we included a single multiplicative factor $m_{DB}$, which modifies the amplitude of both $B_{In}$ and $B_P$ factors to account for possible size effects. In this approximation the structure factor is given by:

$$F_{hkl}(m_{DB}) = \left(f_{In}e^{-m_{DB}B_{In}Q_{hkl}^2} + f_P e^{-m_{DB}B_P Q_{hkl}^2} e^{\frac{i3\pi l}{4}}\right)\left\{1 + exp\left[i2\pi\left(\frac{h+2k}{3} + \frac{l}{2}\right)\right]\right\} \quad (5)$$

where $B_{In} = 0.088\ \mathrm{nm}^2$ e $B_P = 0.057\ \mathrm{nm}^2$ (bulk values); $f_{In}$ e $f_P$ are the atomic scattering factor for the In e P, respectively (Gao & Peng, 1999; Kriegner et al., 2013).

Finally, utilizing Equations 4 & 5 in 2 & 3 we obtain the equation utilized to calculate the diffracted intensities:

$$I_{hkl}^{PED}(A,t,m_{DB},\beta,\rho) = \frac{A\pi^2}{\xi_{hk}^2\,(m_{DB})}\int_0^{2\pi} \frac{sen^2\{\pi t s_{eff}[\xi_{hkl}(m_{DB}),s(\beta,\rho,\theta)]\,\}}{\{\pi s_{eff}[\xi_{hkl}(m_{DB}),s(\beta,\rho,\theta)]\}^2}\,d\theta \quad (6)$$

The integral in the expression is performed numerically and imply in five free parameters in our residue optimization procedure:

1) $A$: scale factor;

2) $m_{DB}$: changes in the Debye-Waller factors;

3) $t$: crystal thickness;

4) $\beta$: crystal rotation (ACOM parameter);

5) $\rho$: crystal rotation (ACOM parameter);

The crystal orientation is derived through a refinement based on the comparison between the measured and calculated intensities using a residual factor defined as: (Palatinus et al., 2015)

$$R_N{}^2 = \frac{1}{N}\,\frac{\sum_N\left(\sqrt{I_{hkl}^m}-\sqrt{I_{hkl}^s}\right)^2}{\sum_N\sqrt{I_{hk}^m}} \quad (7)$$

where $I_{hkl}^m$ is the measured intensity and $I_{hkl}^s$ the simulated one. This residue value is normalized by the number of peaks ($N$) in each diffraction pattern to allow for comparison between EDs with lower and higher values of $N$. In this work, when the residue $R$ is shown in percentual units, the $R_N{}^2$ is multiplied by $N$, such that we can compare the quality of the structural description obtained with previous works in the literature. The addition of $1/N$ in $R$ do not change the minimization parameters for individual EDs, as it is only a scale factor in the residue evaluation.

The number of diffraction peaks (or fitting values) in each ED may be rather low (27 to 7 peaks have been utilized in this study), so we have chosen to further reduce the number of free parameters in our ACOM application. Our highest concern arises for the NW sides, where we have observed a strong thickness variation and significant effect of the amorphous oxide layer. It would be better to avoid any potential overfitting of the measurements due to the low information content over this wire region. In this sense, we have decided to optimize the Debye-Waller scaling factor $m_{DB}$ globally, such that all EDs will have the same value. The thickness $t$ has also been optimized globally, however the shape of the nanowire can be very complex, such that a more in-depth discussion may be need is this case. A cylindrical shape is the most reasonable expectation for the NW shape, however there are reports showing that semiconductor NW surfaces may show prominent faceting (Sun et al., 2015; Wagner et al., 2010). We have used VADF intensity to test which possibility would be the most appropriated for our data; the VADF NW profile does not follow the expected cylindrical shape, suggesting that it is reasonable to consider a faceted NW cross-section and uniform thickness. This has been assumed for the wire region displaying highest SNR and that is suitable for the diffracted intensity analysis (see Supplementary Information section S.4 for more details). This constant thickness assumption may be refined in future works, but the comparison of both models (cylindrical and faceted) has not shown relevant differences in the results for the present study. In fact, one of the strengths of PED is that it its well-recognized approach to diminish the sensibility of the intensities to small changes in sample thickness (Midgley & Eggeman, 2015; Oleynikov et al., 2007). Initially we have attempted to also refine the $A$ value globally, however this was not possible as this demanded all diffractions patterns to be rescaled by the transmitted diffraction beam intensities. This procedure implied in severe discretization of the measured intensities (possible due to numerical truncation of very large numbers during division or by saturation of the transmitted beam intensity), which consequentially discretized all the optimized parameters. Consequentially, the data analysis presented here is based on the optimization of the parameters $A$, $\beta$ and $\rho$ for each ED, such that we can obtain the orientation of each measured pattern of the NW through a residue factor comparing experimental and simulated intensities.

# 3. Results

Figure 2 shows a TEM image of the Eshelby twisted InP NW, where we can observe a thin amorphous oxide layer, and also a contrast variation at the NW axis revealing the existence of the axial screw dislocation. Figure 2b also indicates the region scanned for the 4D-STEM data set (3500 patterns, 50x70 pixels, pixel step 1.59 nm). A quick look at a typical diffraction pattern (Figure 2c) reveals that the crystal orientation is close to the $[2\bar{1}\bar{1}0]$ zone axis and that the misalignment is rather low. Using a simple Laue circle fit on the diffraction spot positions we may verify that zone axis misorientation is in 1-2° range (Edington, 1975).

Figure 3 show diffraction patterns from pixels located such that their positions form a rectangle over the NW in such way that we are able to observe possible crystal orientation differences across de wire center (due to screw dislocation), and also along the wire due to the Eshelby torsion. As expected, the PEDs patterns show significative changes, what can be detected even with a simple visual inspection.

## 3.1 Results of ACOM based on intensity analysis.

We have chosen as our coordinate system the case where the crystal is perfectly aligned along $[2\bar{1}\bar{1}0]$ zone axis along the $x$ axis (anti-parallel to TEM optical axis); the $\boldsymbol{c}$ vector of wurtzite crystal is along the $z$ axis. Concerning crystal rotation analysis, NW torsion induces the basal plane vectors ($\boldsymbol{a}$, $\boldsymbol{b}$, and $\boldsymbol{d}$ axes) to rotate by $\beta$ ($R_z$) and the basal plane normal (hexagonal $\boldsymbol{c}$ axis) will tilt due to the screw dislocation by $\rho$ ($R_y$, see data treatment section and Figure S4b, c).

Figure 4 shows the different steps realized to perform the crystal orientation optimization based on the intensity of the diffraction spots. Firstly, the ED is indexed, and subsequently the Laue circle (see section S.5 in the Supplementary Information) is fitted on spot positions to obtain initial angular values of $\beta$ and $\rho$ for orientation refinement (Figure 4a and b). Then, the free refinement parameters are optimized to minimize the residue ($R_N$, Equation 7) comparing experimental and simulated intensities; an example of the result of this procedure is show in Figure 4c. Overall, the residue values are below 20% (see Figure 4d), such that we can confirm that a good structural description has been obtained for the ensemble of the NW; the residue value is similar to previous reports on structural refinement based on PED pattern analysis using two-beam approximation (Palatinus et al., 2015). The optimized Debye-Waller factors are 0.61 $nm^2$ and 0.40 $nm^2$, for the In and P atoms, respectively. These values are slightly lower than the crystallographic base data for room temperature (In: 0.88 $nm^2$, P: 0.57 $nm^2$), but this can be explained by the increase in relative intensity for high scattering vector $Q$ in PED patterns, which induces an effective reduction of the derived Debye-Waller factor (Gao & Peng, 1999; Oleynikov et al., 2007). The combination of good residues and coherent Debye-Waller factors values ensure that the results obtained are reliable for the majority of the diffraction patterns.

The quantitative analysis of diffracted intensity yields the crystal rotation angles $\beta$ and $\rho$ directly (two free parameters of the residue optimization). To reveal the screw dislocation, we plot the angle $\rho_{dis} = \rho - \rho_{center}$ (value of $\rho_{center}$ is associated to pixels on the axis where the screw dislocation core is located) for each measured position at both sides of the NW (see Figure 5c) (Ugarte et al., 2019). A clear opposite behavior is observed when comparing the NW left and right sides. The Eshelby crystal torsion is observable by the plot of $\beta_{twist} = \beta - \beta_{base}$ (mean value of $\beta_{base}$ across the wire in one of the NW extremities, see Figure 5e). As expected, the basal plane rotation increases along the NW, revealing the crystal torsion around $\boldsymbol{c}$ axis.

# 4. Discussion

As observed in the previous section the ACOM based on PED intensity analysis has easily revealed the helicoidal nature of the basal planes, as well as the induced crystal torsion in the NW crystal (Eshelby, 1953; Eshelby, 1958). Nonetheless, it is important to corroborate these results by comparing them with other well-accepted ACOM methods, specially with the ones that only utilize the diffraction peak position as main source of information. In order to compare the quality of generated results, we can quantify the dispersion of the values of $\rho_{dis}$ and $\beta_{twist}$ (named $\sigma_\rho$ and $\sigma_\beta$, respectively) using the root mean square deviation; this will provide a reliable estimation of the angular precision and dispersion in crystal orientation determination.

The most popular methodology for ACOM is based on pattern matching, this approach has been initially proposed by Nanomegas and also adopted by different available freeware programs (Cautaerts et al., 2022; Ophus et al., 2022; Rauch et al., 2010; Viladot et al., 2013). In this work, we have used the proprietary software ASTAR for the comparison, which has already utilized for the characterization of Eshelby twisted InP NWs (Ugarte et al., 2019). We must note that pattern matching ACOM is capable of identifying the presence of a screw dislocation and measure crystal torsion (Figure 5). The angular precision obtained with ASTAR is much broader (~1$^o$ angular dispersion $\sigma_\rho$ ) what is about one order larger that the intensity analysis approach (see Table 1). This can be due to fact that initially ACOM has been designed to target the analysis of large areas (several $\mu^2$) where crystal grains suffer large and abrupt changes in crystal orientation (> 10$^o$). Usually, the number of PED patterns in the data set is in the several tens of thousands and this kind of crystal orientation study requires to map a large volume space in reciprocal space and with a high speed. The efficient cross-correlation metric has been chosen, but this metric is prone to get stacked in local minima (Cautaerts et al., 2022). Many researchers tried to improve the angular resolution of the pattern matching method with improved instrumentations, image treatment methodologies and better sampling of reciprocal space in simulations, but angular resolution has remained around ~1$^o$ (Cautaerts et al., 2022; Harrison et al., 2022; Jeong et al., 2021; Ophus et al., 2022; Ugarte et al., 2019). The Laue circle approach only relies on simple fit of a circle to the PED spots positions, ignoring intensity of the diffraction spots. It is extremely fast and simple to apply; also, it does not request any kind of simulations library or previously knowledge about the crystalline structure. Nonwithstanding, the extracted information may be rather limited, being mainly appliable to derive the angular distance to the closest crystal zone axis (Edington, 1975). Although its simplicity, the Laue circle method has been able to reveal the presence of the screw dislocation and crystal torsion for the rather small NW region analyzed here (see Figure 5a and c). We can observe in Table 1 that the Laue circle angular dispersion $\sigma_\rho$ is much lower than using the template-matching method, what enable us to observe changes below 1º in crystal orientation (see section 5 in the Supplementary Information for more details). Overall, the Laue circle fit appears to be a good alternative to the common template matching procedure to derive small changes in orientation in a small sample region as required for a strain study. Finally, the derived angular dispersions are almost twice the ones obtained by the intensity analysis proposed in this work (Table 1).

Table 1: Summary of the results obtained from ACOM based on different data treatment strategies (Laue circle, template matching-ASTAR and diffracted intensity analysis).

| Method | Twist rate (°/nm) | $\sigma_\rho$(°) | $\sigma_\beta$ (°) |
|---|---|---|---|
| Laue circle | $0.010 \pm 0.003$ | 0.24 | 0.32 |
| ASTAR | $0.031 \pm 0.008$ | 3.0 | 1.2 |
| Diffracted Intensity | $0.006 \pm 0.001$ | 0.14 | 0.18 |

The torsion angle derived from template-matching (Figure 5e) appears to present two populations; this could be explained by the presence of two close local minima in the template-matching procedure, a tendency which has already been observed (Cautaerts et al., 2022). The Laue circle provided a twist evolution along the wire where several plateaus are clearly observable (Figure 5c), this indicates the lack of precision to detect small PED features. In contrast, when using the diffracted intensity analysis, the twist evolution is mostly continuous with a monotonous positive slope (see Figure 5e). Table 1 shows the comparison of measured twist rates, which measures the wurtzite basal plane rotation due to Eshelby torsion (it is supposed to be constant change rate along the NW due to the short NW section that has been analyzed). The twist rates derived from each ACOM methodology are significatively different (Table 1) and also differ from theoretically calculated value of $(0.0219 \pm 0.0001)°/nm$ (Equation 1, for a cylindrical NW with $(48 \pm 7)\ nm$ in diameter and Burgers vector equal hexagonal axis $c$). The torsion rate evaluations with most precision (Laue circle and intensity analysis) have superimposed values, and both underestimate the theoretical twist rate by at least a factor of 3x. Such deviation could be explained by some factors: i) the NW is possibly faceted and not cylindrical; ii) the NW has a thick oxide layer on its surface (~ 7 nm) which can relax the distortion in the structure. Considering the available data, we have been able to discuss ACOM precision with the intensity analysis, but we are not able to estimate the accuracy. In order to estimate accuracy, we would need a Eshelby twisted nanowire with rotation of more than 30º along the analyzed region to cross two zone axis (*e.g.* $[\bar{1}010]$ and $[\bar{2}110]$, see Figure 3 in (Ugarte et al., 2019)); these zone axes provide a good standard for calibration.

Briefly, the three methods have been able to detect both dislocation and torsion deformation fields in the NW structure (see Figure 5), however a quick visual inspection of the data indicates massive differences in the characteristics of the results. We took great care to ensure that each method was optimized to yield the best results, which required only some minor adaptations between data reduction for each method. The main difference was the threshold values utilized to select valid diffraction spots in the PED pattern (3% and 7% of $I_{0000}$ for, respectively, the Laue circle and intensity analysis). For the ASTAR processing, we utilized a gamma function to enhance the low intensity peaks threshold value (traditional approach in pattern matching procedures), then the threshold cannot be directly compared with the other two approaches. We do not expect that such minor changes have had any major influence in the comparisons between methods. The above discussion shows that intensity analysis provides the crystal orientation and twist values with highest precision among all methods (Table 1).

The exploration of diffracted intensity allows the measurement of disorientations with a precision much better than previous works for the case of deformation fields in nanowires (Ugarte et al., 2019). The obtained angular dispersion has been similar to the reported precision of individual frames orientation during tomography of reciprocal space experiments (Palatinus et al., 2019). These experiment must correct the angular values indicated by TEM goniometer mechanics, and the authors have used a PED diffracted intensity refinement to determine the correct crystal orientation. The intensity analysis procedure is analogous to structure refinement data treatment, such that the residue value is the important parameter to be considered. The residue normalized by the number of peaks (see Figure S9a) values tend to increase close to one of the NW sides (right) in relation to values for in the NW center. This is expected considering the lower quantity of information (less diffraction spots) in such EDs and also the higher relevance of the amorphous oxide shell in relation to the crystalline phase. However, the other NW side, show a different behavior and the normalized residues remain at same tendency than the center. This may relate to a certain asymmetry observed in the virtual images and the number of peaks (see Supplementary Information, Figures S5b, S7b and S9b), possibly related to small differences in thickness in each side of the NW. This indicate that we must be careful in properly deciding the region in which the measured disorientation is reliable. A proper criterion may be determined in the region where the measured twist disorientations start to deviate from the radial neighbor's tendency. No observable separation, different tendency or sudden variation of orientation distribution has been observed in the analyzed region using the diffracted intensity based ACOM (see Figure S8).

Most of our fits have $R_N < 20\%$; this value is considered to be a good quality when analyzing PED patterns by two-beam simulations (Palatinus et al., 2015). This implies that the crystal structure is well described by assumption taken to the optimized free parameters in our simulations: crystal orientation described by 2 angles and a simple rescaling of the Debye factor for In and P. It must be emphasized that the residue values are comparable with ones in previous works even through the precession angle is much lower. Our PED 4D-STEM data have used a precession angle of ~ 0.5° (typical angle for SPED experiments in uncorrected TEM instruments), while previous works targeting electron crystallography and structural refinement have used precession angles above 1° (Ciston et al., 2008; Eggeman et al., 2010; Midgley & Eggeman, 2015; Own et al., 2006; Palatinus et al., 2013). It is our understanding that the careful consideration of the precession geometry by the numerical integration along the excitation error $g_z$ (Equation 4), represent an essential issue for the generated ACOM good quality refinements.

The analyzed Eshelby twisted NW represent a very interesting atomic arrangement due to the superposition of two orthogonal deformation fields. In fact, all pixels should show a crystal on a different orientation. However, there are two very interesting aspects that can be used to precisely estimate the method precision. As the torsion depends only on axial position, the disorientation $\beta - \beta_{center}$ of the basal plane orientation should be constant across the wire (perpendicular to NW axis). In our case, only a small length of the twisted NW has measured (~70 nm), so it is reasonable to think that this wire sector is straight with a constant $c$-axis orientation for pixels on the wire center. (Ugarte et al., 2019). This provides us with a second opportunity to measure a constant optimized parameter ($\rho$) in the sample by observing the orientation of the wurtzite $c$-axis along the NW center. Both the Laue circle and intensity analysis corroborate that no significative tendency is observed for the changes in $\rho$ along the NW axis or center region (see Figure 6). In this case, the precision (dispersion for the values $\rho$ in the NW center) for the Laue circle is ~ 0.06° and for the intensity analysis is ~ 0.03º. This result indicates the high quality of the crystal orientation precision that can be obtained by the diffracted intensity analysis in 4D-STEM using PED.

It is important to corroborate the robustness of our results by observing the consistency of the several steps taken in our data treatment. During optimization, the 2 angles ($\beta$ and $\rho$) describing crystal orientation are the most relevant free parameters for residue minimization in our analysis. A small variation of any of these parameters induce significant residue changes; we have observed that changes of ~ 0.1º are enough to produce a residue difference of ~ 1% (see Supplementary Information, Figure S10). As the precision of orientation values in our study is ~ 0.03°, this implies that it is not possible to differentiate models (crystal in distinct orientations) with $R_N$ differences around ~ 0.2% for our experimental conditions (SNR, maximum scattering angle, *etc.*)

The work discussed in this manuscript uses a quite different approach to evaluate ACOM (diffracted intensity quantification) in which we focus on precision over a rather small dataset (few thousand of PED patterns). The residual factor metric is a very robust and popular approach in crystallography with x-ray, neutron or electron for structural refinement. The residual metric has proved to be suitable even for cases where many local minima are present, although all the process (simulation, parameter variation and comparison) may be rather slow compared to the efficient pattern matching algorithm. Our analysis has taken about ~ 22h to be completed in common desktop computer: Intel i5-8400 CPU (2.8GHz), 32 Gb of RAM DDR4, 1Tb HD, UHD Intel 630 GPU. The ACOM results presented here are based on the analysis of ~ 750 PED patterns; similar crystal orientation optimizations has been reported for ED-based tomography of reciprocal space where 50 - 100 ED patterns constitute the data set.

# 5. Conclusions

Our results show that PED-based ACOM can be significatively improved by exploring quantitative analysis of the intensities to attain an angular precision of ~ 0.03º. This opens a wide range of new possibilities to analyze precisely small crystal orientation changes such as Eshelby twist NW, roughness in bidimensional material, *etc.* (Meyer et al., 2007; Shao et al., 2022; Ugarte et al., 2019). After the crystal orientation optimization procedure, residue values for each pixel were similar to previously reported structural refinement of PED data and two-beam dynamical diffraction analysis (Palatinus et al., 2015). Although, the used precession angle is smaller (~ 0.5º) than conventionally accepted for electron crystallography structural refinement (1 - 3º, *e.g.* electron tomography in (Palatinus et al., 2015)), the high-quality results are most likely due to a proper integration of excitation error for PED provided by the Gjonnes formalism (Gjonnes, 1997).

The homemade developed software has performed the analysis of around 750 diffraction patterns in ~ 22h of processing in desktop computer; but we estimate that this could be improved with more computational power and algorithm optimization. We must note that we have used a two-step approach to accelerate data treatment: 1) firstly, initial crystal orientation values are derived from the rapid and simple Laue circle fit and, 2) optimization based of intensity comparison between model and experiment. We think that a similar approach may be implemented by considering an initial step based on template-matching, but using a low sampling in reciprocal space to build the library.

The study described in this manuscript has focused on the analysis of experimental diffracted intensities to determine crystal orientation and no exploration of the peaks position has been attempted. If we remind that a small shear deformation can be describe as a pure strain plus a rotation, (Kelly & Knowles, 2012) the simultaneous exploration of both characteristics (diffraction disk positions and high precision crystal misorientation) can provide a much more complete characterization of the deformation in nanomaterials by exploring the whole potentiality provided by PED 4D-STEM dataset. Further improvements in data acquisition exploring recent instrumentation developments (direct electrons detectors, aberration correctors, higher precession angles, *etc.*) will certainly lead to further improvements for evaluating crystal distortion with nanometric resolution.

# 6. Acknowledgements

D. U. acknowledges financial support from the Brazilian Agencies FAPESP (No. 2014/01045-0), CNPq (402571/2016-9, 306513/2017-0, 402676/2021-1, 303025/2022-0) and FAEPEX-UNICAMP (2632/17). M. A. C acknowledges financial support from FAPESP (Nos. 15/16611-4 and 19/07616-3) and CNPq (No. 429326/2018-1). A. P. and E. O. gratefully acknowledge the financial support provided by U.S. Department of Defense W911NF1810439. L.M.C. acknowledges financial support from CAPES (no.1765876/2018) and CNPq (no. 140596/2020-8).

# 7. Conflict of Interest

The authors declare no competing financial interest.

# Figures

# Figure 1

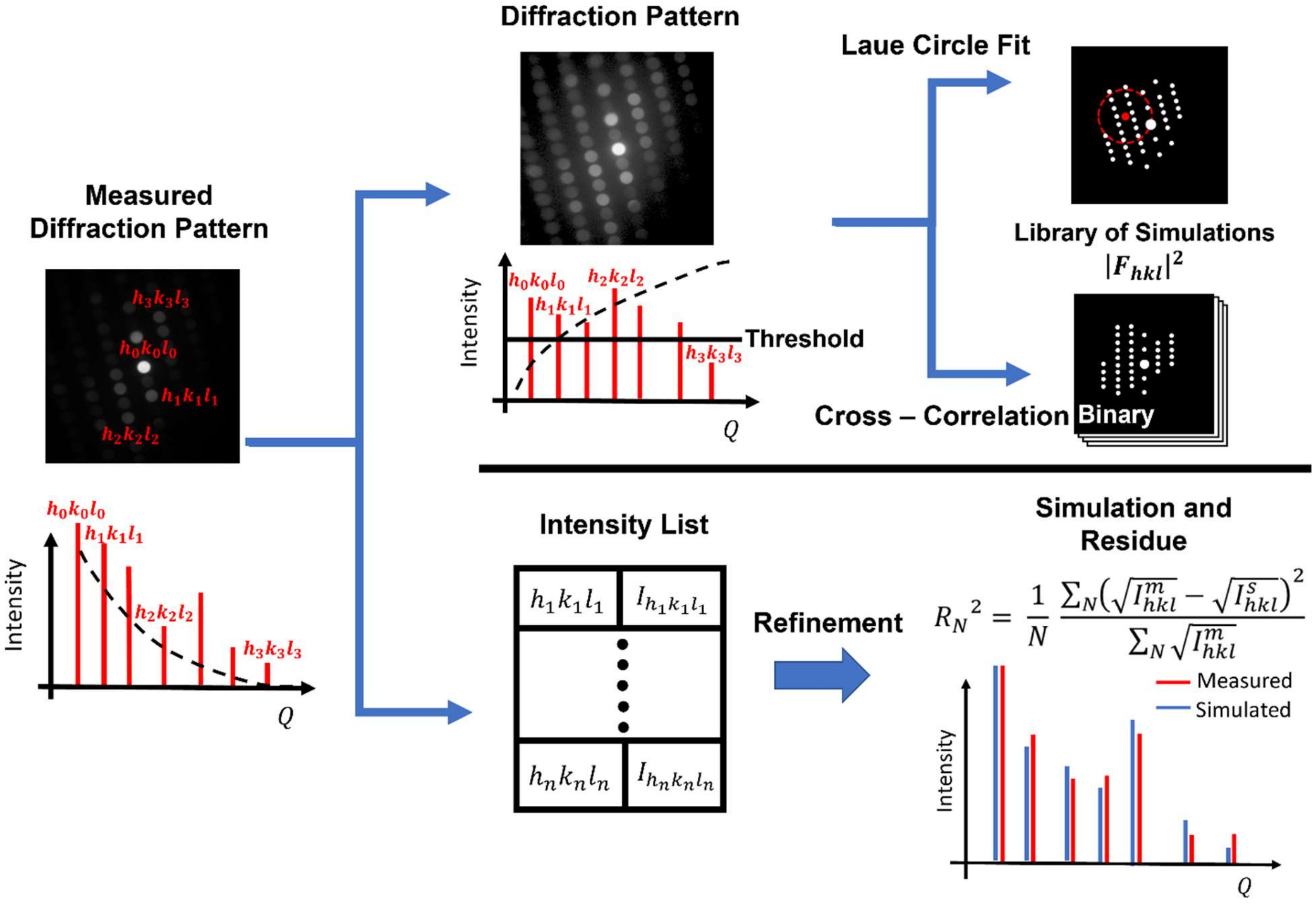

**Figure 1**: Comparison between different ACOM methods. Explanation of differences between the template matching method (threshold and binarization of intensities) and raw intensities use.

# Figure 2

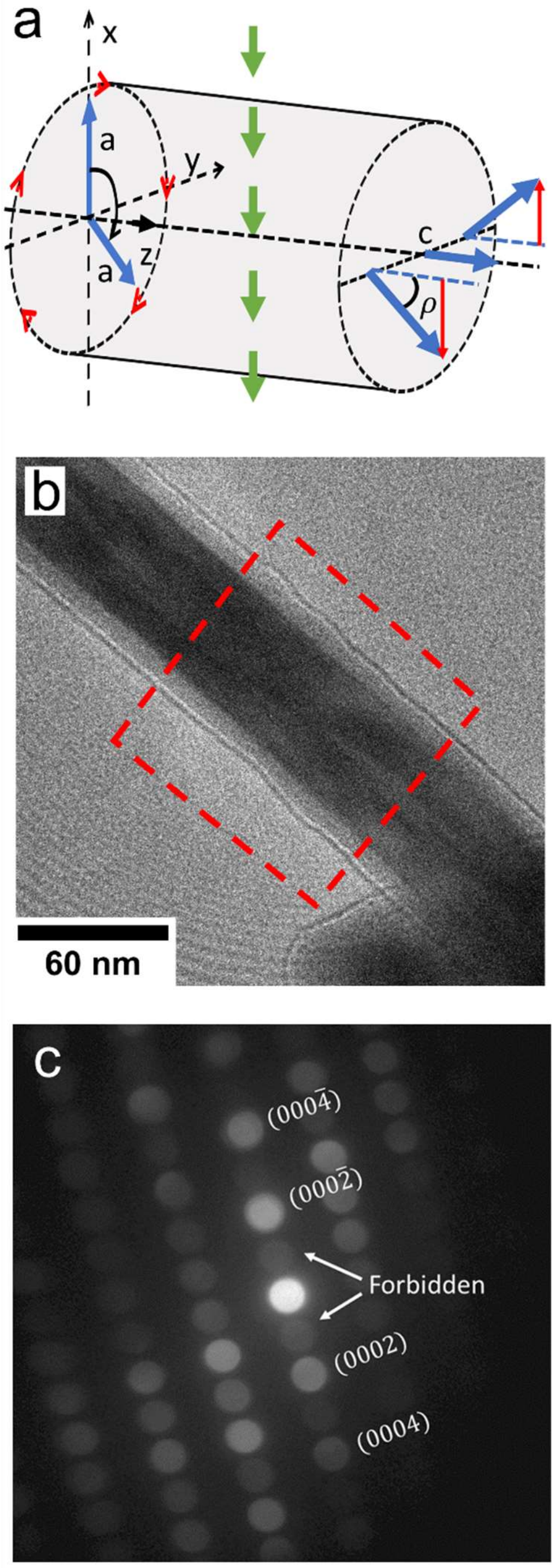

**Figure 2:** a) Schematic draw of the NW crystal orientation changes due to the screw dislocation and Eshelby twist deformation fields (Eshelby, 1953; Eshelby, 1958). b) TEM image of the nanowire; region indicated by the dashed rectangle has been scanned to generate the 4D-STEM-PED dataset. c) Example of a measured PED pattern, some peaks have been indexed along the wurtzite ***c*** – axis direction (or the NW growth direction) to show the presence of low intensity kinematical forbidden diffraction peaks.

# Figure 3

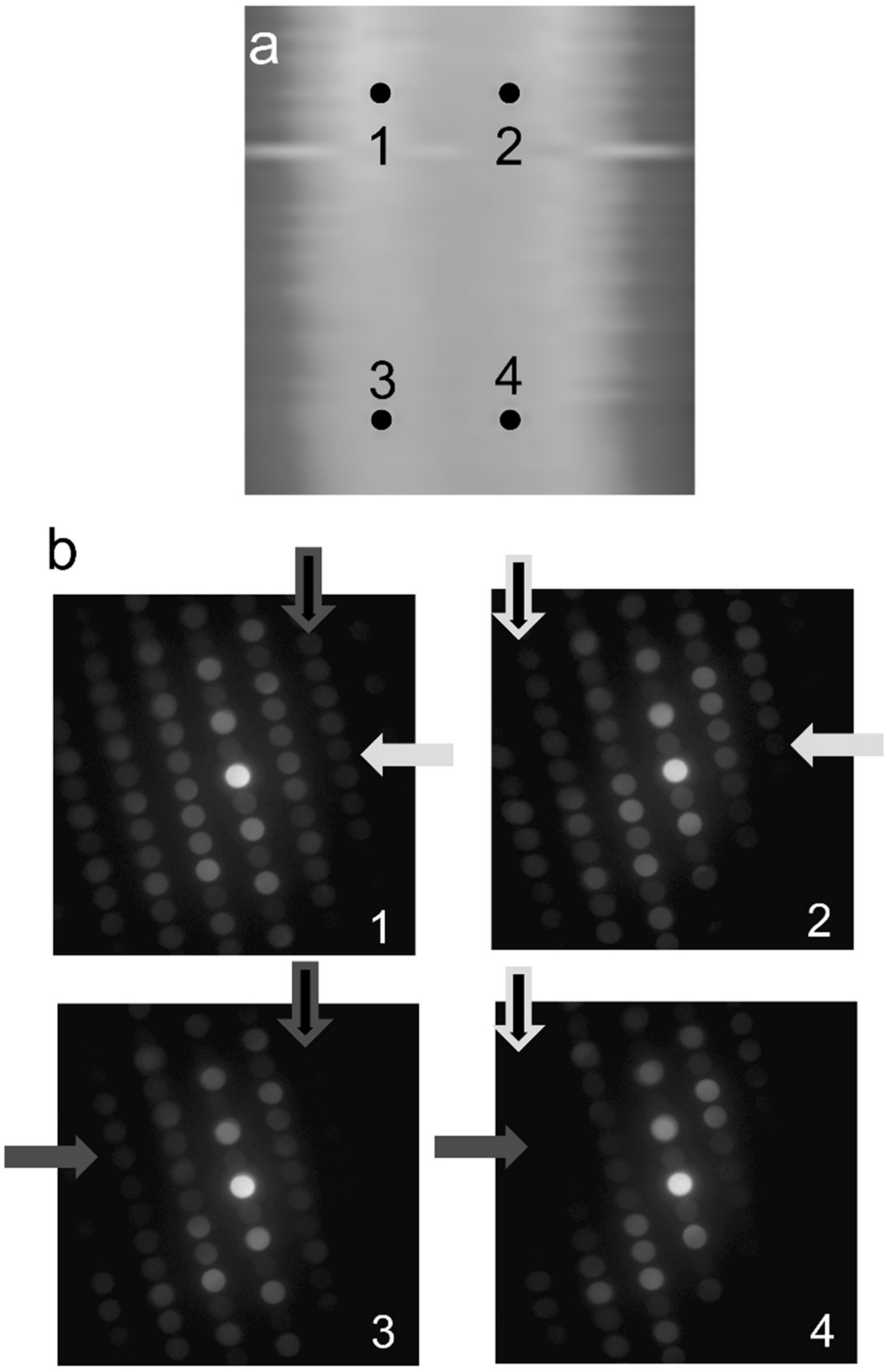

**Figure 3:** Experimental data showing clear differences in PED patterns at different NW position. a) VADF image which shows the 4 distinct pixels in the NW, that are related to the 4 PED patterns below (b). The 4 arrow pairs show peaks or features displaying major distinctions between each ED both from the left to right or top to bottom. Notice that the differences are noticeable both across the NW center (where the screw dislocation is located) and along the NW length (due to Eshelby crystal torsion). The contrast of the displayed pattern images have been corrected to aid the visual analysis (gamma: 1.76, contrast: -18 and brightness: -29, IrfanView software).

# Figure 4

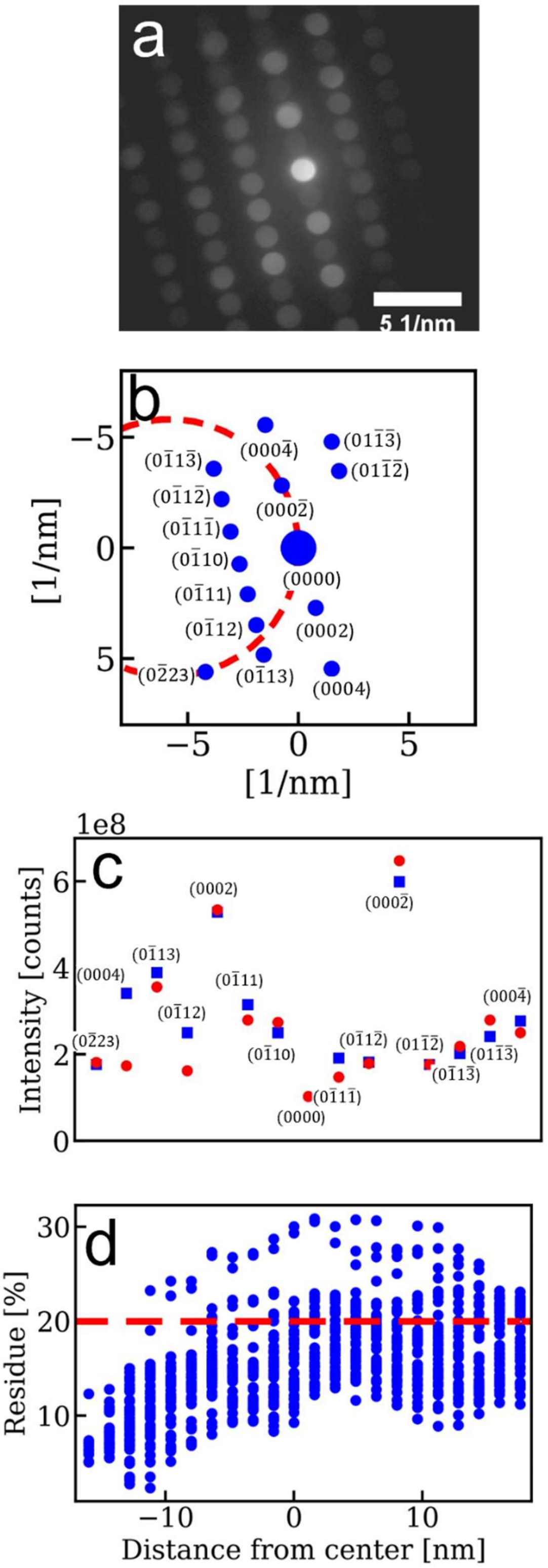

**Figure 4:** Example of the intensity analysis procedure for: a) single pixel (the PED pattern). b) Indexed PED pattern, where $hkil$ index have been retrieved, as well as intensity $I_{hkl}$ (peak selection is based on a minimal intensity threshold of 7% from $I_{0000}$); diffraction peak positions are utilized to fit a Laue circle in order to provide the initial of $\beta$ and $\rho$ for the crystal orientation optimization. c) Measured intensities (squares) and optimized simulation (circles) derived from the best model derived by residue factor minimization. d) Residue values for all the measured pixel positions along the NW (or PED patterns) in our analysis; the dashed line marks the 20% residue value, which is considered a good quality factor for two-beam PED patterns refinement (Palatinus et al., 2015).

# **Figure 5**

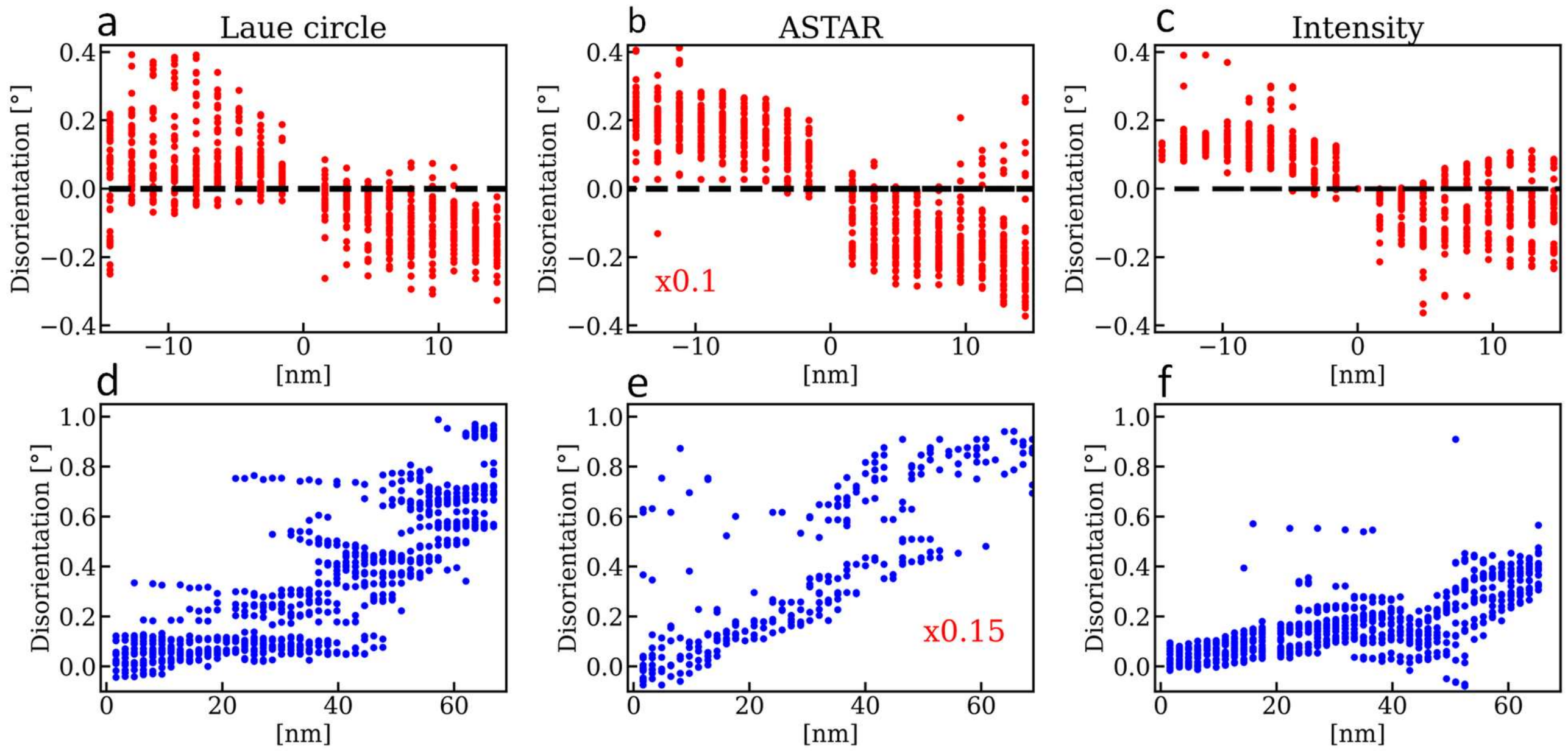

**Figure 5:** Comparison between the results of dislocation deformation field (a,b,c) and Eshelby torsion (d,e,f) for the ACOM methods.

# **Figure 6**

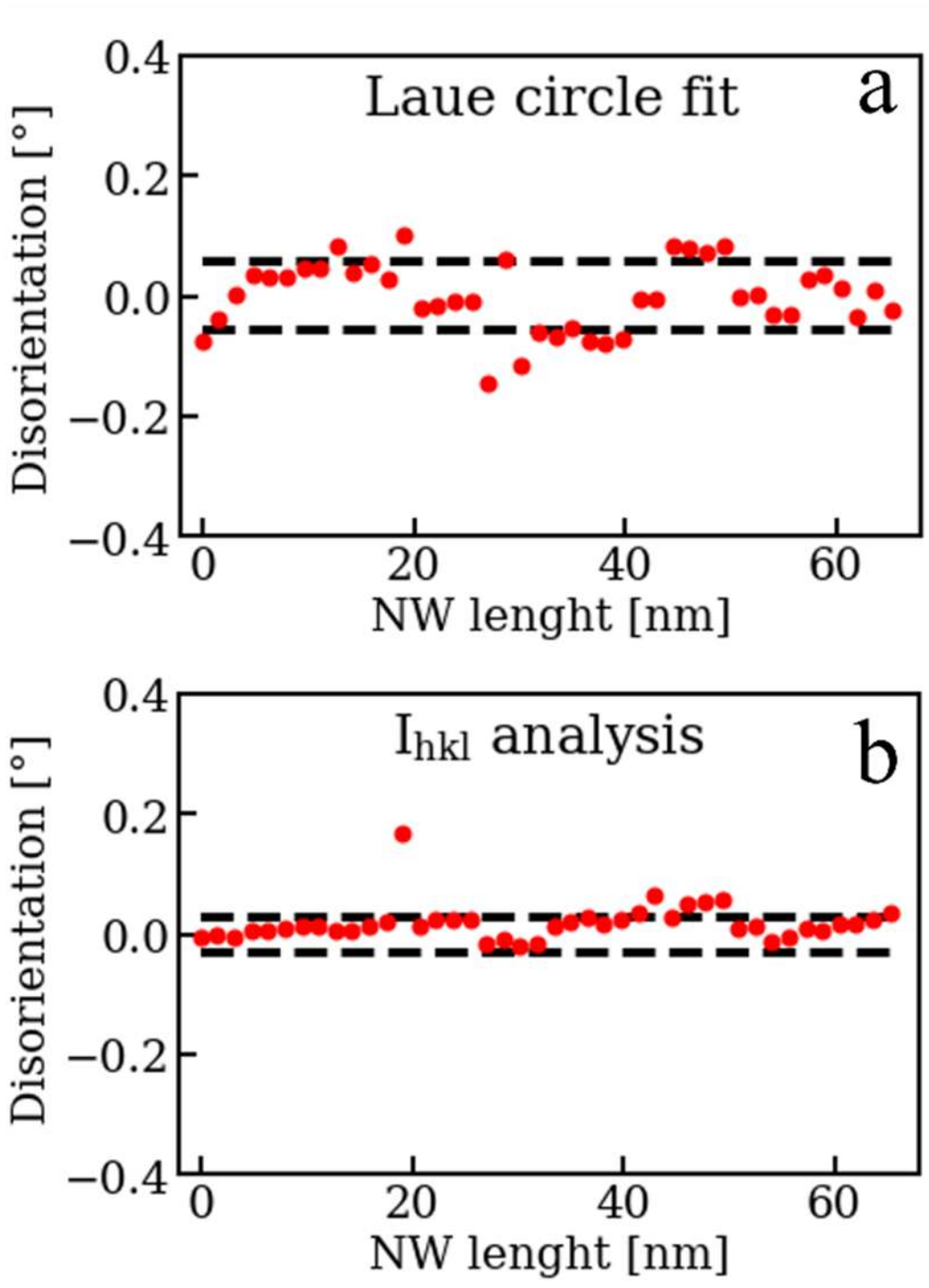

**Figure 6:** Disorientation (in relation to the mean value) of $\rho$ (hexagonal $c$-axis orientation) at the NW center derived from a) the Laue circle fit and, b) intensity optimization. The scales have been chosen such that the precision can be compared with the dispersions in Figure 5a-c.

# Supplementary Information

## High Precision Orientation Mapping from 4D-STEM Precession Electron Diffraction data through Quantitative Analysis of Diffracted Intensities

**Leonardo M. Côrrea[1], Eduardo Ortega[2], Arturo Ponce[2], Mônica A. Cotta[1], Daniel Ugarte[1*]**

1. Instituto de Fisica "Gleb Wataghin", Universidade Estadual de Campinas-UNICAMP, 13083-859, Campinas - SP, Brazil
2. Department of Physics and Astronomy, University of Texas, San Antonio, Texas 78249, United States

## S1. Assembly of the PED-based 4D-STEM data set

The original spatially resolved data has been acquired in a detection system which was not prepared to assign the precession electron diffraction (PED) pattern and the corresponding spatial location on the nanowire (NW). The original data was acquired as movie on CMOS axial camera while PED attachment performed the precession and scanning of the electron beam. To convert the diffraction pattern series into 4D data, we have calculated virtual annular dark field images (VADF) for each pattern. The VADF pixel intensity is formed by integrating the intensity of a selected annular region of the electron diffraction pattern (ED) which contains only diffracted peaks (excluding the central peak or transmitted beam), as show in Figure S1a. The center of each cross-section of the NW will represent the center of the peaks in the VADF intensity constructed as a movie. To determine the centers, we utilized the cross-correlation function between the formed VADF profile and an ideal wire profile (for simplicity we utilized a top-hat function, see Figure S1b). The profiles were then aligned to construct the 4D datablock (see Figure S1c), such that during the analysis is clear that the NW center is set along the datablock center. It is important to notice that this procedure is only possible due to previous knowledge about the region scanned by the electron beam, as seen in Figure 2b.

## S2. Diffraction peak identification, intensity integration, indexation and calibration

The diffraction disks from each ED need to have its center determined and intensity integrated in order to exploit diffracted intensity values. To solve such issue, we have utilized the cross-correlation between each ED and an ideal diffraction disk (top-hat type, binary form, 1 inside the disk, 0 outside it). The diameter has been determined from experimental patterns, representation very closely measured diffraction spots, such that both have the same convergence angle (see Figure S2a). After the application of cross-correlation, a new ED is obtained where the intensities in the center of each diffraction disk are maximum, and the maxima have the value of the integrated spot intensity in the original pattern. The cross-correlation also smooths the diffraction peak profile, as show in see Figure S2b, such that we can easily find the disk centers by searching local maxima in the cross-correlation image (Figure S2c). This procedure is similar to the ones used by other software's of ACOM, but the properties of intensity integration is usually not explored, as the methodologies do not explicitly aim to explore intensity values (Johnstone et al., 2021; Savitzky et al., 2021).

Each peak need to be indexed to be utilized in our analysis, such that we have included an automatic peak indexation algorithm in our analysis procedure. As all diffraction patterns measured are near the $[2\bar{1}\bar{1}0]$, we need only to find:

$$h = \frac{\boldsymbol{g} \cdot \boldsymbol{v}^*}{|\boldsymbol{v}^*|^2}, k = \frac{\boldsymbol{g} \cdot \boldsymbol{c}^*}{|\boldsymbol{c}^*|^2}, i = -(i + k), l = 0 \quad (S1)$$

where $\boldsymbol{c}^*$ and $\boldsymbol{v}^* = \boldsymbol{b}^*/cos(30°)$ are the reciprocal vectors related to the $\boldsymbol{c}$ and $\boldsymbol{b}$ lattice vectors of the hexagonal primitive, $\boldsymbol{g}$ is the scattering vector for each identified peak. The $\boldsymbol{g}$ vector is obtained by multiplying the identified centers by the proper scale factor (obtained from the manual indexation and calibration of one the measured EDs).

## S.3 Details of the Laue circle analysis.

The orientation of a crystal severely affects the observed diffraction patterns, as it alters the observed projection of reciprocal space which intercepts the Ewald sphere (known as the Laue circle). A slight tilt of the crystal away from a zone axis can significatively move the position of the center of Laue circle ($O_L$) from the position of the transmitted beam ($T$) (usually considered the diffraction pattern central spot, see Figure S3). It is possible to measure the resulting disorientation ($\phi$) from the zone axis direction by determining the radius of the Laue circle ($R_L$), as they are directly related by: (Edington, 1975)

$$R_L \lambda = \sin(\phi) \quad (S.2)$$

where $\lambda$ is the wavelength of the incident electron wave. This simple relation can be easily implemented in procedures for crystal orientation determination or alignment, as it only requires a simple fit of a circle to the diffraction peaks position (Ben-Moshe et al., 2021; Zhang et al., 2018; Zhang et al., 2020). This procedure does not require any assumptions about the crystal structure or previous information about the sample. In our case, we can derive the angles $\rho$ and $\beta$ (see Figure S4) from the fit of the Laue circle to the binarized version of the measured diffraction patterns. This is possible as angles $\rho$ (tilting of $c$ axis in relation to the plane perpendicular to the optical axis of the TEM) and $\beta$ (basal plane rotation) are related to the [0001] and [hki0] directions, respectively, which are identified by indexing the diffraction pattern.

For analysis, the indexed ED has been binarized by a threshold value that exclude any diffraction peak with intensity lower than 3% of the direct beam intensity. Then, a simple circle fit is performed, in which the circle equation utilized is given by:

$$(x - R_0 \cos\theta)^2 + (y - R_0 \sin\theta)^2 = {R_0}^2 \quad (S3)$$

where $R_0$ and $\theta$ are free parameters that minimize the distance between the circle and the diffraction points. It is possible to use Equation S.2 to obtain the angles $\beta$ and $\rho$ if the related lengths to cartesian axes ($L_\beta$ and $L_\rho$) are determined. This can be performed by projecting the vector formed by $T$ and $O_L = (R_0 \cos\theta, R_0 \sin\theta)$ into the $\boldsymbol{c}^*$ and $\boldsymbol{v}^*$ directions, as show in Figure S3. From that we can obtain $\beta$ and $\rho$ through:

$$\begin{cases} \sin(\beta) = L_\beta \lambda \\ \sin(\rho) = L_\rho \lambda \end{cases} \quad (S4)$$

## S.4 Nanowire shape and size analysis.

As described in the main text, the formation of virtual VADF images is fundamental for our analysis and is important to understand the contained spatial information with the maximum possible precision. We have decided to explore VADF images, as the detection has been optimized to ensure the quality of the diffracted peaks, while the central beam is possibly saturated and variations in it could not be related to structural features (a virtual bright field, VBF). All VADF have been generated by integrating the intensities in an annular region with 6 mrad of inner half angular opening and 23 mrad in the outer one (see Figure S5a).

The NW shape analysis has been performed in the integrated cross-section of the formed VADF (summation of all the cross-sections along the NW length), this allows an easier interpretation and better signal to noise ratio (*e.g.* in Figure S5b the error bars due to Poisson noise are smaller the points dimensions). We have tested the possibility of a cylindrical NW by calculating the VADF intensity ($I^{VADF}$) as a function of distance to the NW center considering a convolution between the circular profile of the NW and gaussian profile of the electron beam which scans the sample: (Carter & Willians, 2009).

$$I^{VADF}(x) = C \left[ d \sqrt{1 - \left(\frac{2x}{d}\right)^2} * \frac{1}{\sigma\sqrt{2\pi}} e^{-\left(\frac{x-\mu}{\sqrt{2}\sigma}\right)^2} \right] \quad (S5)$$

where $d$ is the NW diameter, $C$ is a scale factor, $\sigma$ and $\mu$ are, respectively, the width and center of the gaussian. We can see in Figure S5b that the fit of the above expression on the measured intensity profile do not properly describe the NW VADF intensities. This is evident if we superimpose the useful region in our analysis (from a residue quality analysis, maximum distance of 15 nm to the NW center, see Figure S5c). In this region, the root mean square deviation (RMS) of a constant profile (mean value in the region) gives much lower value (0.45) than the RMS of the circular profile (0.67). This suggest the presence of faceting in the NW surface, but, as this is not a rigorous analysis of the NW shape, we will constrain ourselves to state that the measured diffracted intensities are much better described by a constant thickness than by a circular cross-section.

One of the most prominent sources of error in our analysis comes from the determination of the NW crystalline core diameter. This is due to the common imaging methods utilized (as VADF and VBF) lacking sensibility to differentiate abrupt changes from the crystalline core to the amorphous oxide layer in the NW surface (Kiss et al., 2016; Ugarte et al., 2019). An advantage of 4D-STEM techniques is the capacity of explore the measured EDs to generate different types of contrast. One alternative is to utilize the correlation or anti-correlation functions to generate contrast due to changes in the measured EDs (Kiss et al., 2016). This is especially useful to identify the position of change between amorphous and crystalline regions in the NW. We have utilized the most common definition of anti-correlation utilized in ACOM, which compares a diffraction pattern $p_{x,y}$ ($x, y$: spatial coordinate in the ED mapping) with its nearest neighbors ($p_{x+1,y}$ and $p_{x,y+1}$) to form contrast $C(x, y)$: (Kiss et al., 2016)

$$C(x,y) = \sqrt{\frac{\sum_{i,j}\left[p_{x,y}(i,j) - p_{x+1,y}(i,j)\right]^2 + \left[p_{x,y}(i,j) - p_{x,y+1}(i,j)\right]^2}{2n}} \quad (S.6)$$

where $i, j$ is the coordinate of a pixel of the measured ED, white $n$ total pixels in each one. Notice that the definition of $C(x, y)$ is anti-correlation function, although it is often referred as a correlation function (Kiss et al., 2016). The value of $C(x, y)$ is low for similar diffractions patterns and high for very distinct patterns. The correlation function (similar diffraction patterns give high value) is obtained utilizing the inverse value (inverse of contrast) of $C(x, y)$ in each pixel, as defined by (Kiss et al., 2016).

To construct such anticorrelation images some additional treatment is required, as it has been reported that those images are significatively sensitive to ED background intensity (Kiss et al., 2016). The procedure is very similar to the ones performed in other works in the literature, so only a quick description is given and we focus our attention in the quality and interpretation of our correlation images (Kiss et al., 2016). An example of our data treatment is given in Figure S6; first, the central peak is excluded by a simple circular mask in the center of our data, the mask is a little larger that the central beam diameter to ensure that no possible blooming-like effects are included. Then, we create a mask to mitigate the intensity outside the diffracted peaks by applying a local threshold method to our data. An example of the effective mask and the resulting effect in the ED are shown in Figure S6b,c.

We can observe the anti-correlation image in Figure S7a. Most of the NW center appears to have an almost homogenous value, with a gradual maximization of $C(x, y)$ (anti-correlation) in its borders and then a region of low contrast in the vacuum region (diffraction patterns are very similar to each other without diffraction peaks). The integration of the cross-sections in the image along the NW (Figure S7b) shows two peaks in the anti-correlation, which can be interpreted as the region with maximum difference between the crystalline part and the amorphous surface layer in the NW. The peak width is a transition region between crystal and amorphous and its width can be utilized to determine an uncertainty for the crystalline core diameter. From those criteria, we have obtained the NW diameter as $(48 \pm 7)nm$.

## S.5 Results from the Laue circle analysis

The Laue circle approach only requires the proper indexation of each ED and a simple circle fit, from that we can obtain the angles $\beta$ and $\rho$ associate to crystal tilt for each scanned position. A low threshold value was utilized as the method requires many diffraction spots for proper fit. The Laue circle allows the characterization of a larger area of the NW (until 20 nm from NW center) than the more precise intensity residue analysis. This also implies in the inclusion of noisier data, which may lead to lower reliability. The Laue circle derived information is clearly sensible to the dislocation and torsion (see Figure 5) in the NWs, and the results indicate a high sensibility to small values of disorientation. The values of dispersion obtained are much lower than the ones in previous pattern matching reports (~ 1º) (Ugarte et al., 2019). The disorientation $\beta - \beta_{center}$ (basal plane rotation) is shown in Figure S8, with no observable tendency variation until ~15 nm from center. This is expected as the basal plane orientation should be constant along the NW cross section perpendicular to wire axis, such that any variation is only due to noise.

## S.6 Pattern matching Results (ASTAR)

The pattern matching calculation required the simulation of a large library of diffraction patterns (~50000 patterns) to cover the small changes in crystal orientation. The template-matching ASTAR procedure yields 3 Euler angles for each diffraction pattern and additional calculation is required to obtain the disorientations (see (Ugarte et al., 2019) for more details). It is clear that the ASTAR is sensible to both the core dislocation and the torsion, but with varying quality in the observation of each structural feature. The dislocation varies continually withing a large useful area (within 16 nm from the NW center), with only a few points diverging from data point clouds. The torsion behaves in a discontinuous manner, with two clear distinct distributions that present very different torsion rates values ($(0.031 \pm 0.008)°/\mathrm{nm}$ and $(0.067 \pm 0.006)°/\mathrm{nm}$). Only a small fraction of the NW was useful for the torsion calculation, and the locations of the related PEDs present a random behavior all over the wire. This implies that the torsion seen to be much more prone to data noise, with the two populations observed being two distinct local minima distributions in the template-matching cross-correlation procedure. This last behavior has already been observed by other researchers and it is possibly due to different assumption that ASTAR utilizes in data treatment and template simulation (Cautaerts et al., 2022). The dispersion in angular disorientation (see Table 1) seen to be higher than the ones observed in previous works; this could be a consequence of our use of the ASTAR software with data acquired outside the Nanomegas suite (optical camera) for ACOM. The system is extremely well optimized to be utilized together with the non-axial optical camera, and deviations from the characteristics of the expected data (well defined narrow diffraction spots, some distortion, a gamma function applied during acquisition, *etc.*) can imply in the ASTAR software having trouble to properly process our data. Previous studies of InP nanowire based on Nanomegas ASTAR, have taken great care (*e.g.*, data denoising, scan in reduced regions for better signal, more pixels in detection) to obtain the best possible data to measure those small changes in orientation (Ugarte et al., 2019). Those steps are not reproduced in the analysis presented in this paper. Nonetheless, the analysis shows the main limitations and serves for comparison with other techniques, as is performed in the main text.

## S.7 Intensity analysis: consistency and sensitivity

It is important to corroborate the robustness of our results by observing the consistency. This has been tested for torsion by looking for sudden variation of basal plane disorientation in relation to the value in the NW center (see Figure S8). No strong variations have been detected using the intensity analysis. The residue normalized by the number of peaks (see Figure S9a) increases at one of NW sides (right) in relation to values for in the NW center. This is expected considering the lower information (less diffraction disks, see Figure S9a) in the EDs and also the lower signal-to-noise ratio due to the amorphous oxide shell. This residue asymmetry may be associated to thickness differences at each NW side. The analysis of residues indicate that a change in orientation of ~ 0.1° induce a residue variation of approximately 1% . Considering our angular precision of ~ 0.03°, the induced $R_N$ changes would be 0.2% (see Fig. S10).

## S.8 References

# Figure S1

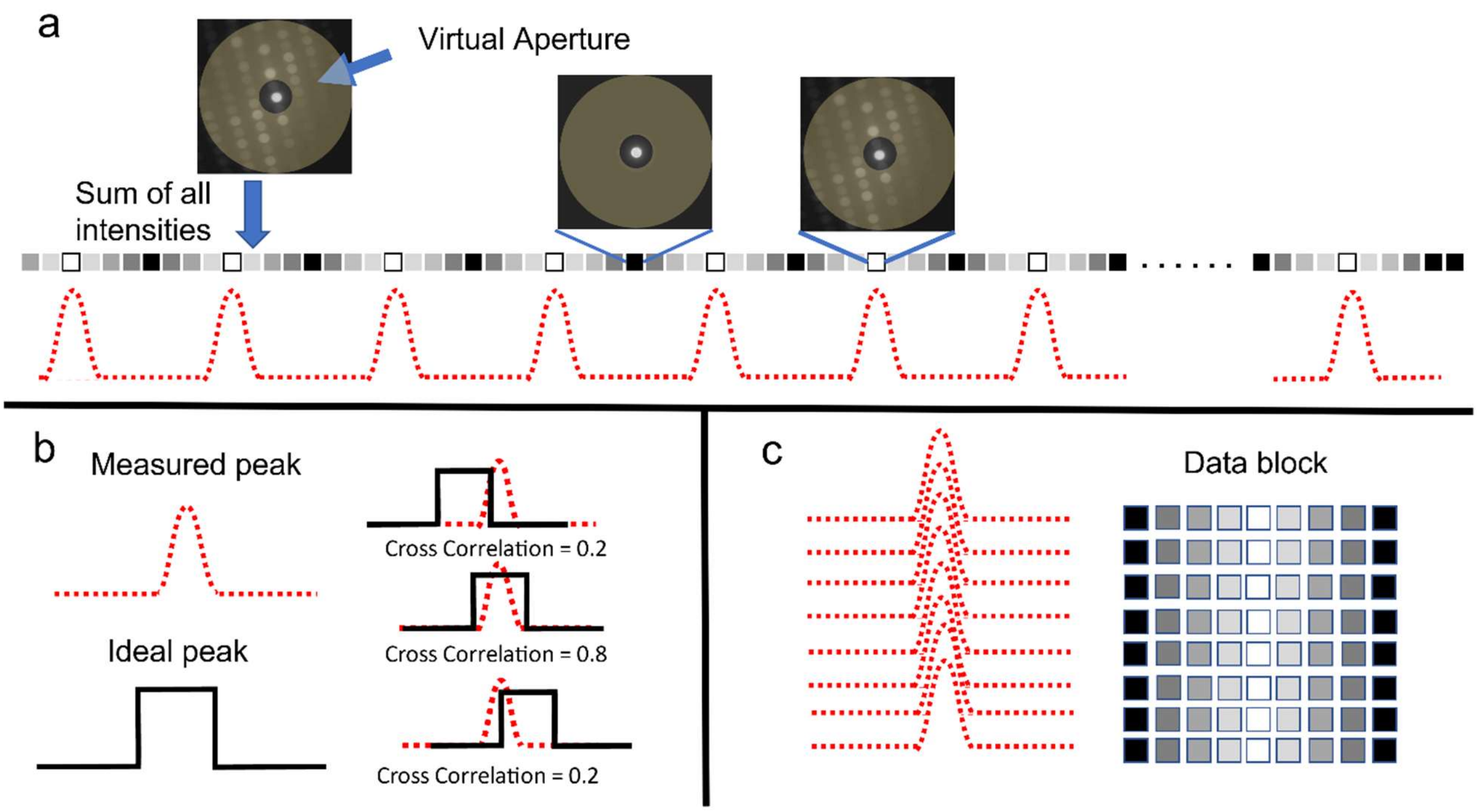

**Figure S1**: Schematic representation of the datablock construction from recorded PED patterns (time based). a) A virtual annular aperture selects a region in the PED pattern in which all the intensities are summed up. This integration forms a sequence of peaks in relation to time, where each peak center is related to the center of the NW position. b) The cross-correlation value between the measured peaks and an ideal peak (top-hat function) is maximum when the center of the two curves are matched. c) Then we can align the centers (and consequentially all pixels) to produce the data block.

# Figure S2

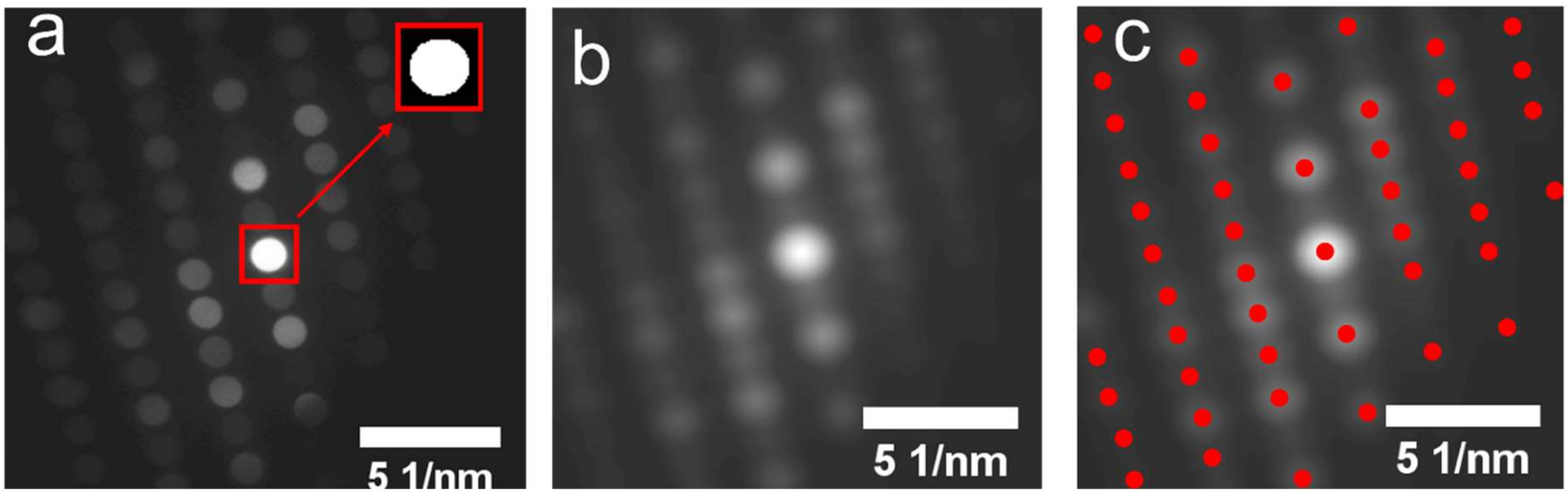

**Figure S2**: Example of the peak finding algorithm. a) The binarized diffraction disk corresponding to the transmitted beam is selected as an ideal peak. b) The result of the cross-correlation between the data and the ideal disk selected in (a). Notice the blurred aspect of the pattern, consequence of the smoothing effect of the cross-correlation. c) Diffraction spot centers identified by a local maxima algorithm applied to the cross-correlation image.

# Figure S3

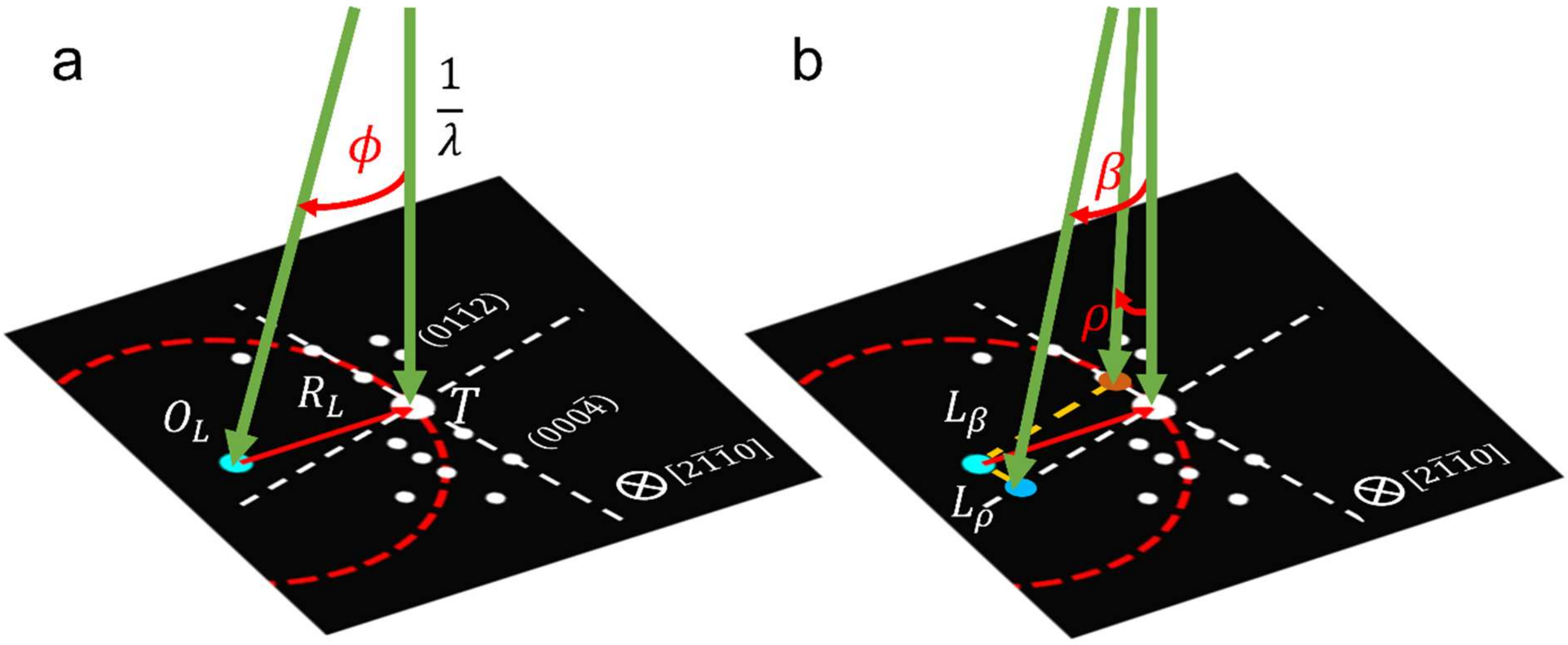

**Figure S3**: Application of the Laue circle fit to a diffraction pattern. a) Effect of a disorientation $\phi$ of the crystal in relation to a zone axis. b) It is possible to determine the disorientation in relation to a specific direction by projecting the $R_L$ vector along those directions.

# Figure S4

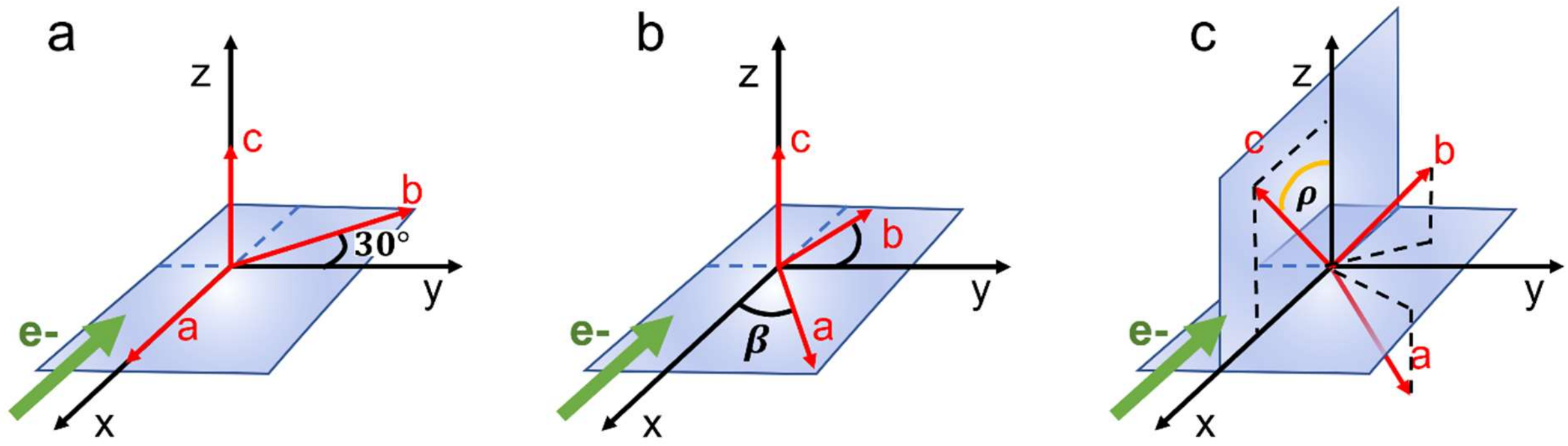

**Figure S4**: Crystal orientation system in relation to sample coordinates. a) Initially, the crystal is perfectly oriented along the microscope optical axis (represented by the static electron beam direction, considered along $x$ axis). b) Rotation of basal plane (torsion) of the NW disorientates the crystal in relation to TEM optical axis ($x$-axis). c) The hexagonal $c$-axis inclination originated by the screw dislocation deformation field generates a disorientation in relation to the plane ($yz$ plane) perpendicular to the TEM optical axis.

# Figure S5

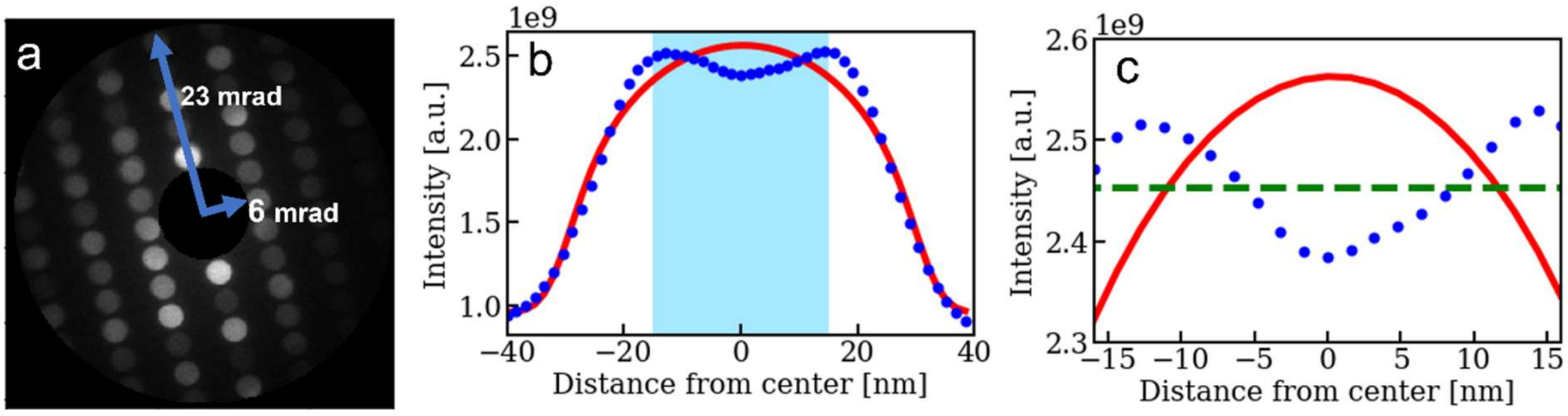

**Figure S5:** a) Annular region values utilized to form the VADF images. b) VADF intensity profile formed by integration of all the cross-sections profiles along the nanowire length (blue dots) and optimized circular cross-section (continuous red line). The highlighted region identifies the high quality (PED patterns) used for our orientation determination by diffracted intensity analysis. c) Closer look at the highlighted region, showing the measured (blue dots), circular cross-section NW (red line) and flat cross-sections (value used in intensity optimization dashed green line) profiles. The error bars are much smaller than the points due to the large number of counts provided by the integration (more than $10^9$ counts).

# Figure S6

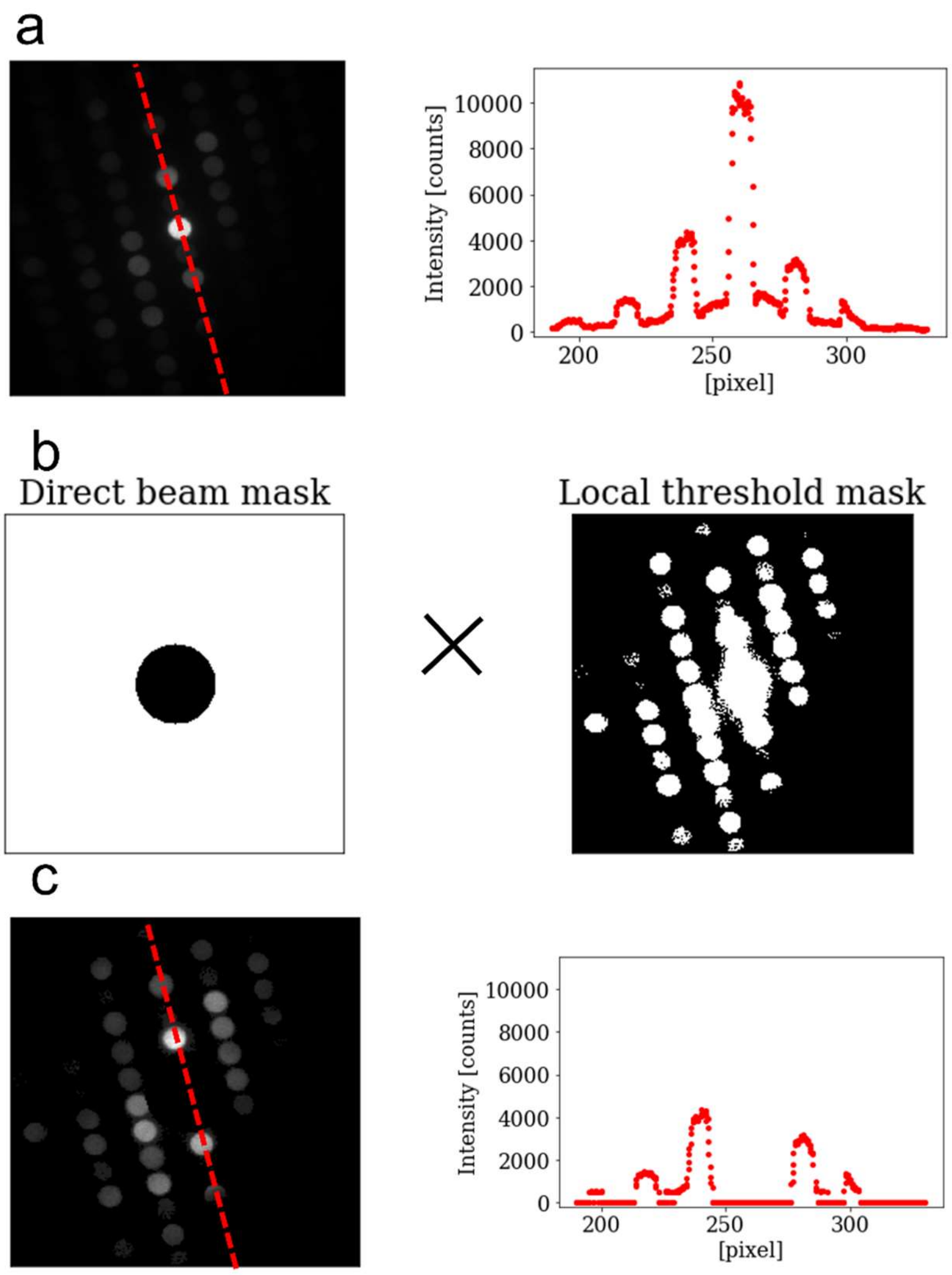

**Figure S6**: Example of the procedure used for maximizing the quality of the cross-correlation images utilized for NW diameter estimation. a) Raw data (left panel) shows significative influence from the central beam scattering and background; intensity profile along the [000l] direction (right panel). b) Circular mask (left panel) utilized to exclude the transmitted diffraction disk in VADF images c) Resulting mask selects only the regions inside the diffraction disks (left panel), which greatly reduces the influence of the background and transmitted beam in the resulting profile (right panel).

# Figure S7

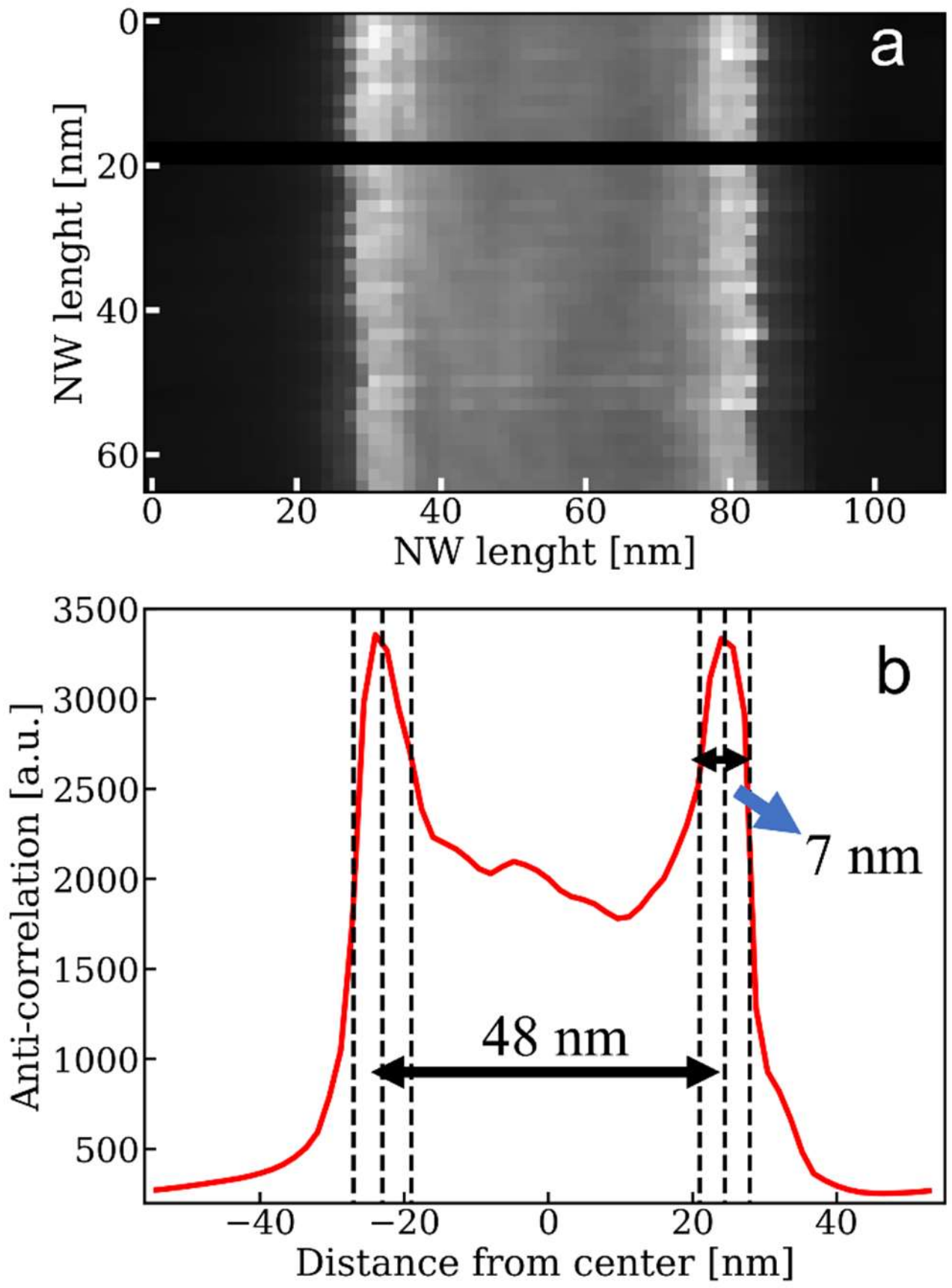

**Figure S7**: a) Anti cross-correlation image of the NW. Notice that a small region just before 20 nm from the NW center is excluded from image formation due to a scan instability during measurement. b) The anti-correlation profile formed by the integration of the cross-section along the NW length. The center of each extremity peak is considered the NW edge, the full width in half maximum is considered the uncertainty of the NW edge position.

# Figure S8

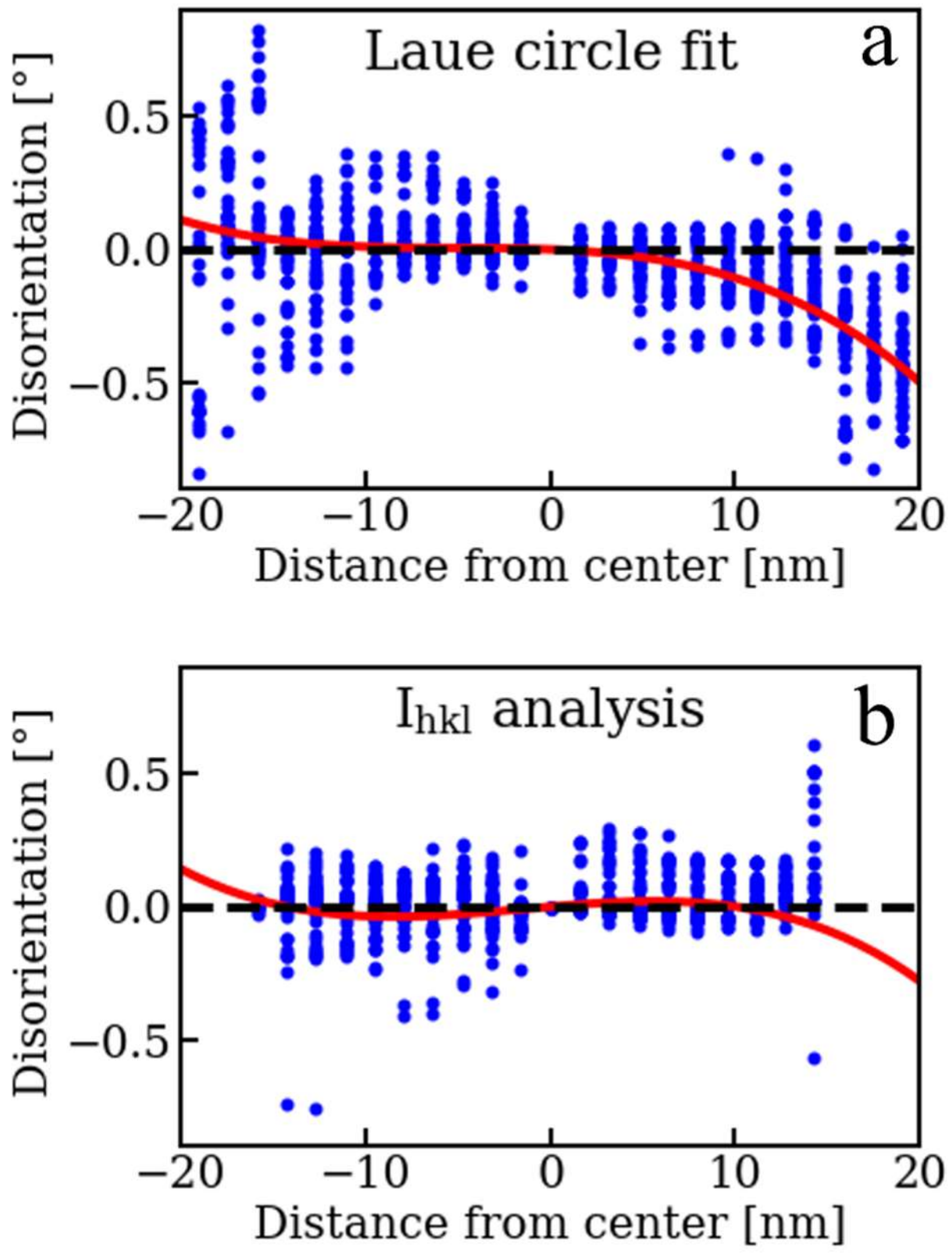

**Figure S8**: Basal plane vectors disorientation (points) in relation to the value at the NW center (data corresponding to all crossing lines is displayed): a) Laue circle analysis and b) diffracted intensity analysis. The polynomial fit (3rd order polynomial, continuous line) helps to observe the tendencies in the data. Notice that a flat curve should be observed in relation to the zero disorientation (see Discussion section in the main text for more details).

# Figure S9

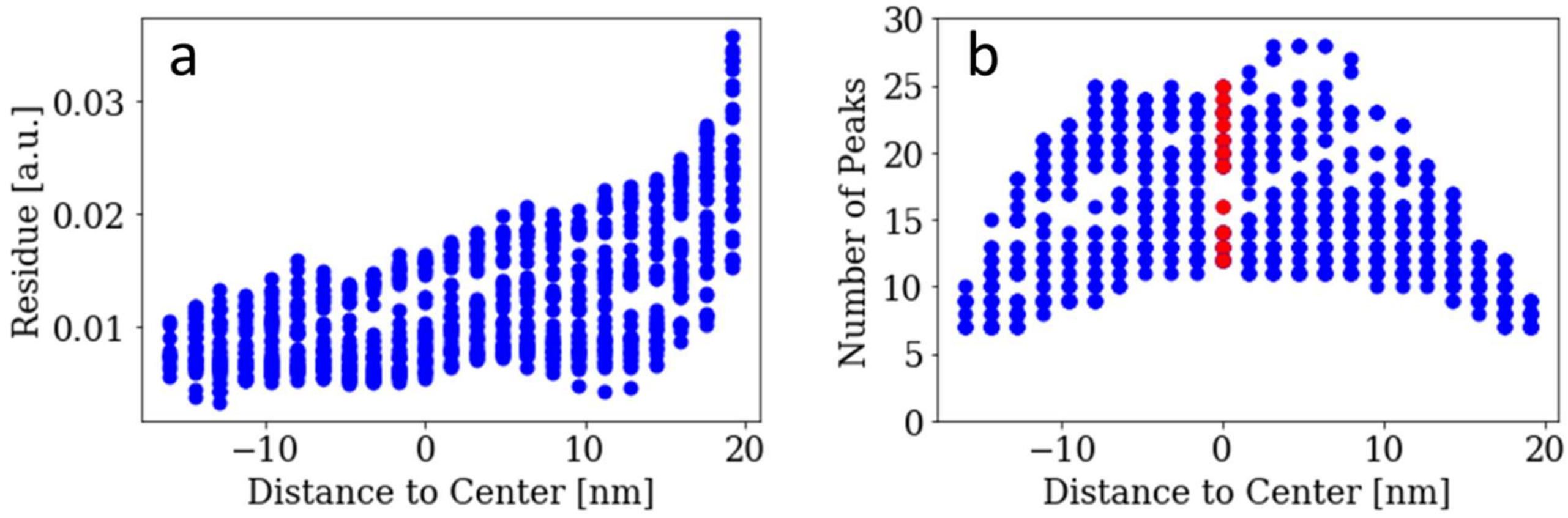

**Figure S9**: a) Values of the Residue normalized by the number of peaks in the PED pattern (all analyzed points over the wire). Notice that an increase of the value in the extremities of the NW would be expected. b) Number of peaks analyzed on the PED patterns.

# Figure S10

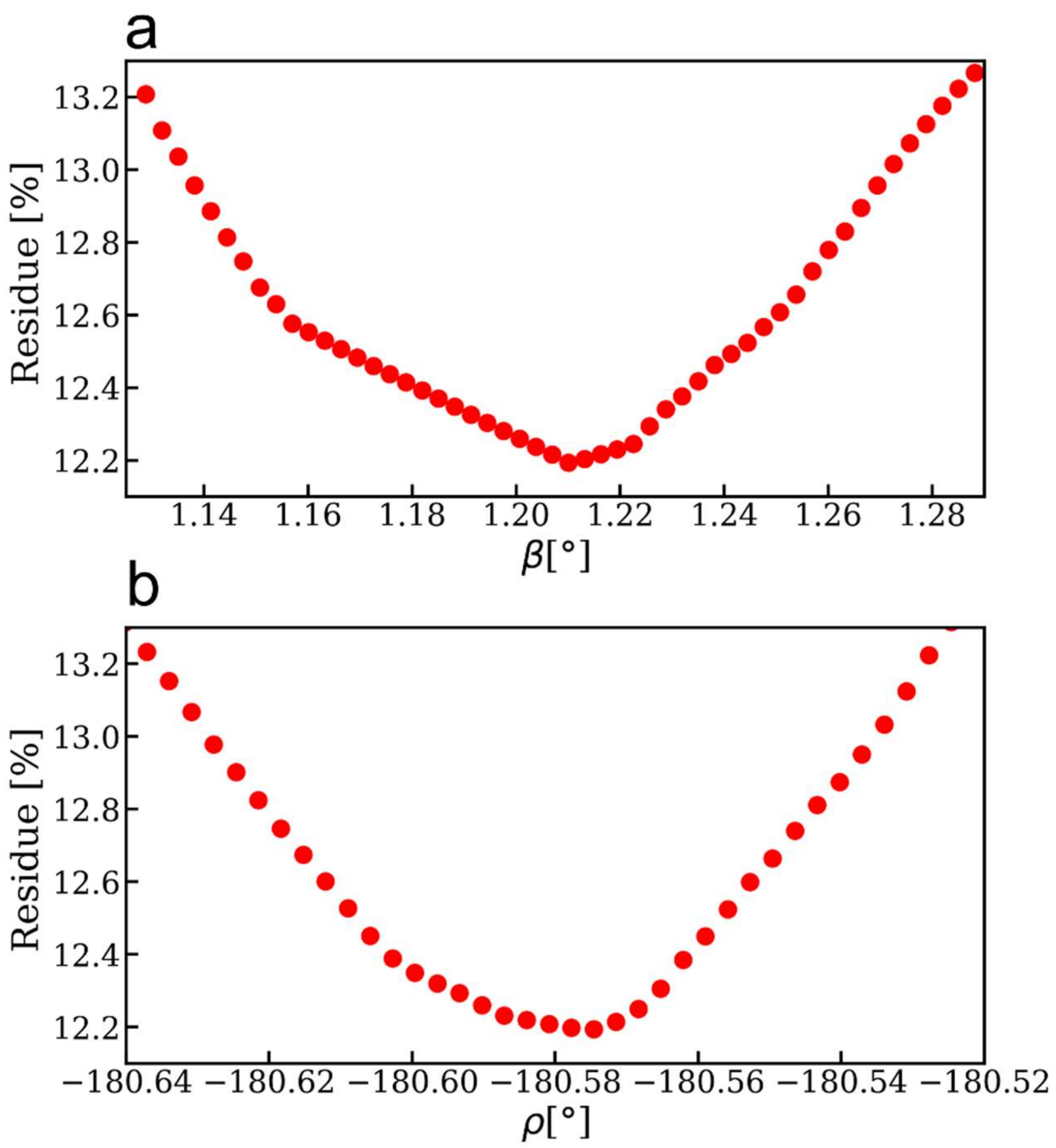

**Figure S10**: Analysis of the sensibility of the residue to small changes in the orientation. a) basal plane and b) hexagonal ***c*** vector (dislocation). In this work, we have considered that only residue differences above 0.2% are significative, as our precision is ~ 0.03°.